\documentclass[pre,aps,twocolumn,
pdflatex,
superscriptaddress,floats,showpacs,10pt]{revtex4-1}
\usepackage{lipsum}
\usepackage{graphicx}
\usepackage{dcolumn}
\usepackage{bm}
\usepackage{amssymb}
\usepackage{amsmath}
\usepackage{textcomp}
\usepackage{color}

\newcommand{\sciexp}[2]{{#1}\ensuremath{\,\times\,10^{#2}}}
\newcommand{\tsp}{\hspace{3mm}}

\begin{document}

\title{Regimes of magnetic reconnection in colliding laser-produced magnetized plasma bubbles}

\author{K. V. Lezhnin}
\affiliation{Department of Astrophysical Sciences, Princeton University, Princeton, New Jersey 08544, USA}
\affiliation{National Research Nuclear University MEPhI, 115409, Moscow, Russia}

\author{W. Fox}
\affiliation{Department of Astrophysical Sciences, Princeton University, Princeton, New Jersey 08544, USA}
\affiliation{Princeton Plasma Physics Laboratory, Princeton, New Jersey 08543, USA}

\author{J. Matteucci}
\affiliation{Department of Astrophysical Sciences, Princeton University, Princeton, New Jersey 08544, USA}

\author{D. B. Schaeffer}
\affiliation{Department of Astrophysical Sciences, Princeton University, Princeton, New Jersey 08544, USA}

\author{A. Bhattacharjee}
\affiliation{Department of Astrophysical Sciences, Princeton University, Princeton, New Jersey 08544, USA}
\affiliation{Princeton Plasma Physics Laboratory, Princeton, New Jersey 08543, USA}

\author{M. J. Rosenberg}
\affiliation{Laboratory for Laser Energetics, University of Rochester, Rochester, New York 14623, USA}

\author{K. Germaschewski}
\affiliation{Department of Physics and Space Science Center, University of New Hampshire, Durham, New Hampshire 03824, USA}

\date{\today}

\begin{abstract}
We conduct a multiparametric study of driven magnetic reconnection relevant to recent experiments on colliding magnetized laser produced plasmas using particle-in-cell simulations. Varying the background plasma density, plasma resistivity, and plasma bubble geometry, the 2D simulations demonstrate a rich variety of reconnection behavior and show the coupling between magnetic reconnection and the global hydrodynamical evolution of the system. We consider both the collision between two radially expanding bubbles where reconnection is seeded by the pre-existing X-point, and the collision between two flows in a quasi-1D geometry with initially anti-parallel fields where reconnection must be initiated by the tearing instability. In both geometries, at a baseline case of low-collisionality and low background density, the current sheet is strongly compressed to below scale of the ion-skin-depth scale, and rapid, multi-plasmoid reconnection results. Increasing the plasma resistivity, we observe a collisional slow-down of reconnection and stabilization of plasmoid instability for Lundquist numbers less than approximately $S \sim 10^3$. Secondly, increasing the background plasma density modifies the compressibility of the plasma and can also slow-down or even prevent reconnection, even in completely collisionless regimes, by preventing the current sheet from thinning down to the scale of the ion-skin depth. These results have implications for understanding recent and future experiments, and signatures for these processes for proton-radiography diagnostics of these experiments are discussed.

\bigskip

\end{abstract}

\maketitle

\section{Introduction}

Magnetic reconnection is a ubiqutous process through which magnetic fields change their topology while releasing magnetic energy into fast particles and plasma heating \cite{REVIEW1, REVIEW2}. This process is important for understanding various phenomena in both astrophysical and laboratory plasmas. For example, magnetic reconnection is thought to be the primary process which triggers solar flares \cite{FLARE} and is believed to be a reason for gamma and X-ray radiation in the magnetosphere of pulsars \cite{CERUTTI}. Similarly, in laboratory fusion plasmas, magnetic reconnection is believed to act as a trigger for sawtooth crashes \cite{SAWTOOTH}.

Recent experiments with expanding laser-produced plasmas have provided a new platform to observe magnetic reconnection in the laboratory \cite{HEDP1,HEDP2}. By focusing powerful terawatt laser pulses onto two spots on a thin target, supersonically expanding plasma bubbles with self-generated (via the Biermann effect, see \cite{Matteucci2017}) megagauss-scale magnetic fields can be formed. If the separation between two bubbles is small enough, the bubbles collide, driving magnetized ribbons of plasma towards reconnection. Similar experimental setups were implemented at a number of facilities such as Vulcan \cite{VULCAN}, OMEGA \cite{Rosenberg2015,RosenbergNatComm,OMEGA2,OMEGA3}, and Shenguang-II \cite{Shenguang} with a variety of driving parameters (such as plasma $\beta$), system sizes ($L/d_i$), and collisionalities (Lundquist numbers, S), showing a wide range of reconnection behavior.

Due to the large system size (in terms of $L / d_i$), laser plasma experiments offer opportunities to study interaction of plasmoid instabilities with dynamically forming current sheets. 
Recent works have shown that for the case of Lundquist numbers $S>10^3$, initially thin current sheets are subject to tearing instability, leading to plasmoid formation, as has been shown in both particle-in-cell (PIC) \cite{DAUGTHON2009} and resistive MHD simulation \cite{Bhattacharjee2009}. Those simulations were initialized with an already-thin current sheet, following common practice. 
On the  other hand,  in the case of the reconnection experiments with magnetized plasma bubbles, the current sheet has yet to be formed. Recent theoretical findings \cite{Pucci2014, Comisso2017,Huang2017} verify the importance of the current sheet formation dynamics, arguing that in astrophysical environments, reconnecting current sheets should break up before they can reach the aspect ratio predicted by Sweet-Parker model. 
While most experiments mentioned above report on fast and complete reconnection, in Ref. \cite{Rosenberg2015} the experiment demonstrated that stagnation of the reconnecting process is possible, which was interpreted as an increase of the collisionality in the current sheet. Thus, we may expect that driving parameters influence the current  sheet  formation  process,  which,  in  its  turn,  affects the reconnection itself. 

In this paper, using fully kinetic 2D PIC simulations with the Plasma Simulation Code (PSC)\, \cite{GERMASCH}, we investigate how plasmoid instability and fast reconnection are controlled by the parameters of the colliding plasmas, in particular the role of collisionality and background plasma density. This is tested for two geometries of spherical colliding bubbles and parallel sheets which mimic very large current sheet lengths (from tens to hundreds of ion skin depths)
, where reconnection must onset due to the pre-imposed X point geometry and tearing instability, respectively {(though, the Y-points of long current sheets can still contain an X-point)}. A base case of low plasma collisionality and low background density was documented in Ref. \cite{HEDP1} which observed fast reconnection and breakup of the dynamically forming current sheet into several smaller scale current sheets by plasmoid instability. Here, we study this system over a range of parameters and observe a rich variety of reconnection behavior.  We identify two important parameters and physics processes which can control the rate of reconnection and energy conversion in the system.

 First, through variation of the plasma collisionality (which is a free parameter in the simulation, parameterized by the Lundquist number $S$), we can observe transitions from a collisionless to collisional regime,  with a corresponding slow-down compared to the collisionless case. Collisionality is well known in other reconnection contexts to mediate plasmoid instability, so these results are valuable as they demonstrate this physics in the regime of HEDP, and demonstrate that it may be studied by parameter scans in HED plasmas. In our simulations, reconnection does not slow down as much as expected from a naive application of Sweet-Parker theory, because compressibility allows the current sheet to be thinner than a SP sheet. We identify two possible mechanisms that slow down the reconnection: 1) the local increase of the collisonality in the current sheet and 2)  the fluid rebound of the two magnetized ribbons. 
 
 Second, we also study the criteria for onset of reconnection in these rapidly-forming current sheets. In addition to collisionality, the background plasma density between the bubbles is found to play a very important role, because it modifies the compressibility of the plasma and controls the thickness (measured in units of the local ion skin depth) of the compressed current sheet. We find that the onset of reconnection is closely tied to obtaining a thin current sheet. If reconnection does not onset, the magnetic ribbons can ``bounce'', which will completely halt the reconnection. The criterion for the onset is found to be $\delta / d_i < 1$, where $\delta$ is the current sheet width and $d_i$ is the local ion skin depth \cite{Amitava2004}. We also find that there is a new time scale characteristic of colliding and bouncing plasmas, $L/c_s$ (length of the current sheet to the sound speed), which is a ballistic time of the bubble expansion. If sufficient tearing growth does not occur on this timescale, no reconnection results. We show below that pure resisitive tearing is almost always too slow to cause a reconnection onset in these systems, but that two-fluid effects or collisionless tearing may lead to sufficient growth rates. This leads again to requirements on the thinness of the current sheet (in units of $d_i$) for a reconnection onset.

Finally, these results are specialized and we make some comments about recent experiments Ref. \cite{Rosenberg2015} which observed a slow-down or stagnation of reconnection. We have attempted to match the parameters of these experiments, and find that for parameters close to reported, the simulations obtain a fast reconnections with multiple plasmoids, and whose signature in the proton radiography is much different than observed, provided the background density is low enough. This leads to the hypothesis that the background plasma density or a similar effect of insufficient compression of the current sheet to the ion scale could be the mechanism to cause the stagnation in these experiments.

The paper is organized as follows. In Section \ref{sec:setup}, we describe the simulation parameters for both bubble and parallel collision cases. In Section \ref{sec:results}, we discuss the simulation results, addressing the observed effects. In Section \ref{sec:summary}, we summarize the results obtained and formulate the main conclusions which may be useful for diagnostic purposes in experimental studies of driven magnetic reconnection.

\section{Simulations setup} \label{sec:setup}
In order to investigate the processes that occur during the driven reconnection experiments, we conduct a series of 2D PIC simulations. The simulations track a pair of expanding bubbles through their interaction and reconnection, which is driven by the substantial energy stored in the plasma flow. 2D cylindrical DRACO and LASNEX radiation-hydrodynamics simulations were conducted to predict the plasma plume evolution and obtain profiles to initialize the PSC reconnection simulations presented here. {These profiles were previously used for simulations of OMEGA experiments presented in Refs. \cite{RosenbergPoP2015,RosenbergNatComm}.}

The profiles were obtained by taking a radial cut in the horizontal plane at $z = 445 \, \rm \mu m$ above the target surface.  DRACO simulations were used to obtain density $n$, temperature $T$, and velocity $V$ profiles, and a parallel set of LASNEX simulations was used to obtain a representative $B$ field profile \cite{OMEGA2} along the same cut, which was near the maximum magnetic fields of the LASNEX simulations.  The cuts were taken at 0.6 ns which is a time after the B fields are generated but before two plumes initially separated by 1.4 $\rm mm$ have collided.

\begin{figure}
    \includegraphics[width=0.95\linewidth]{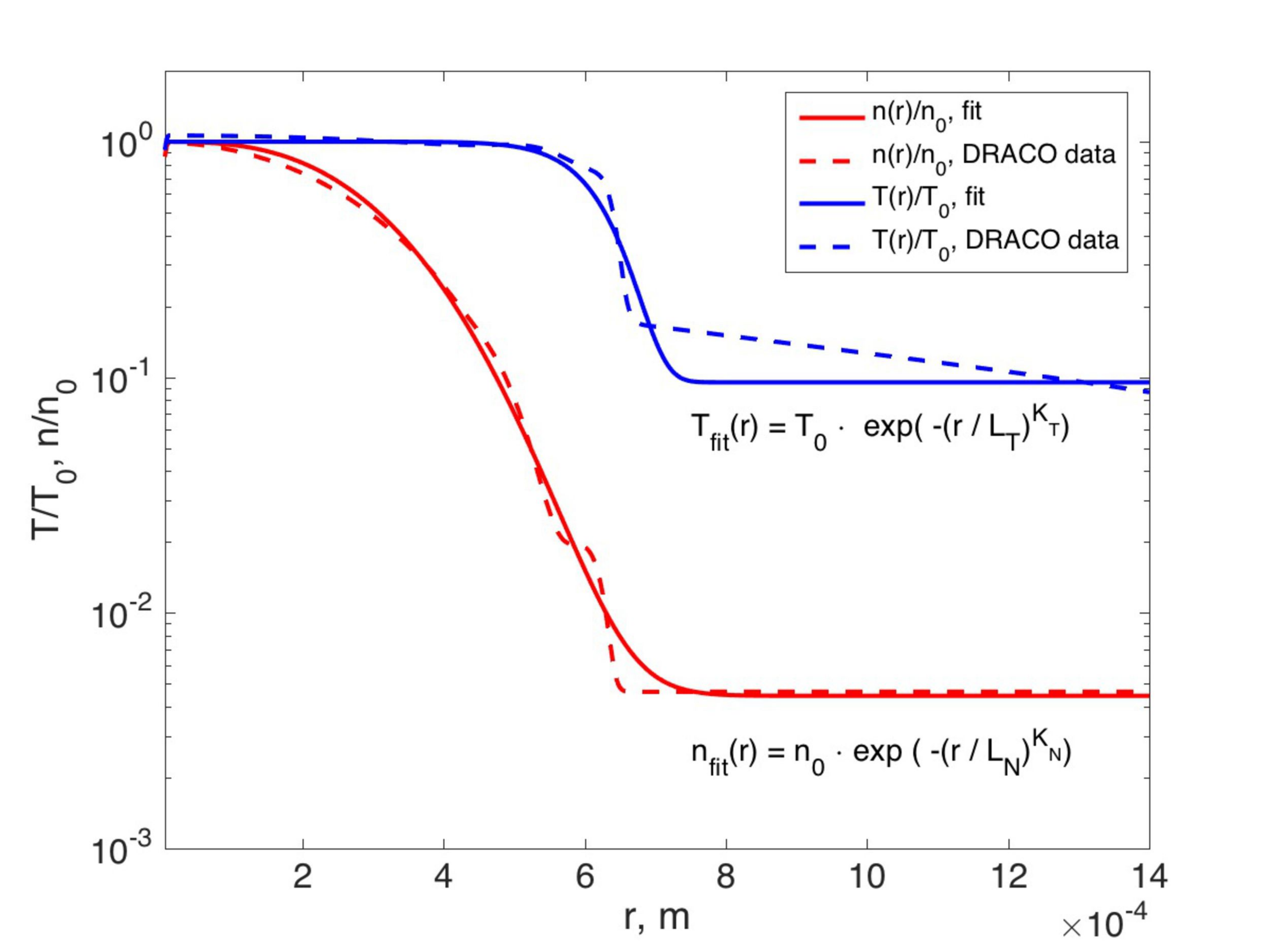}
\caption{Initial radial profile from DRACO for plasma density (red) and temperature (blue). Actual DRACO data is shown in dashed lines, solid lines represent resulting fits in form of supergaussian function $\exp (- x^K)$.}
\label{fig:init1}
\end{figure}

\begin{figure}
    \includegraphics[width=0.95\linewidth]{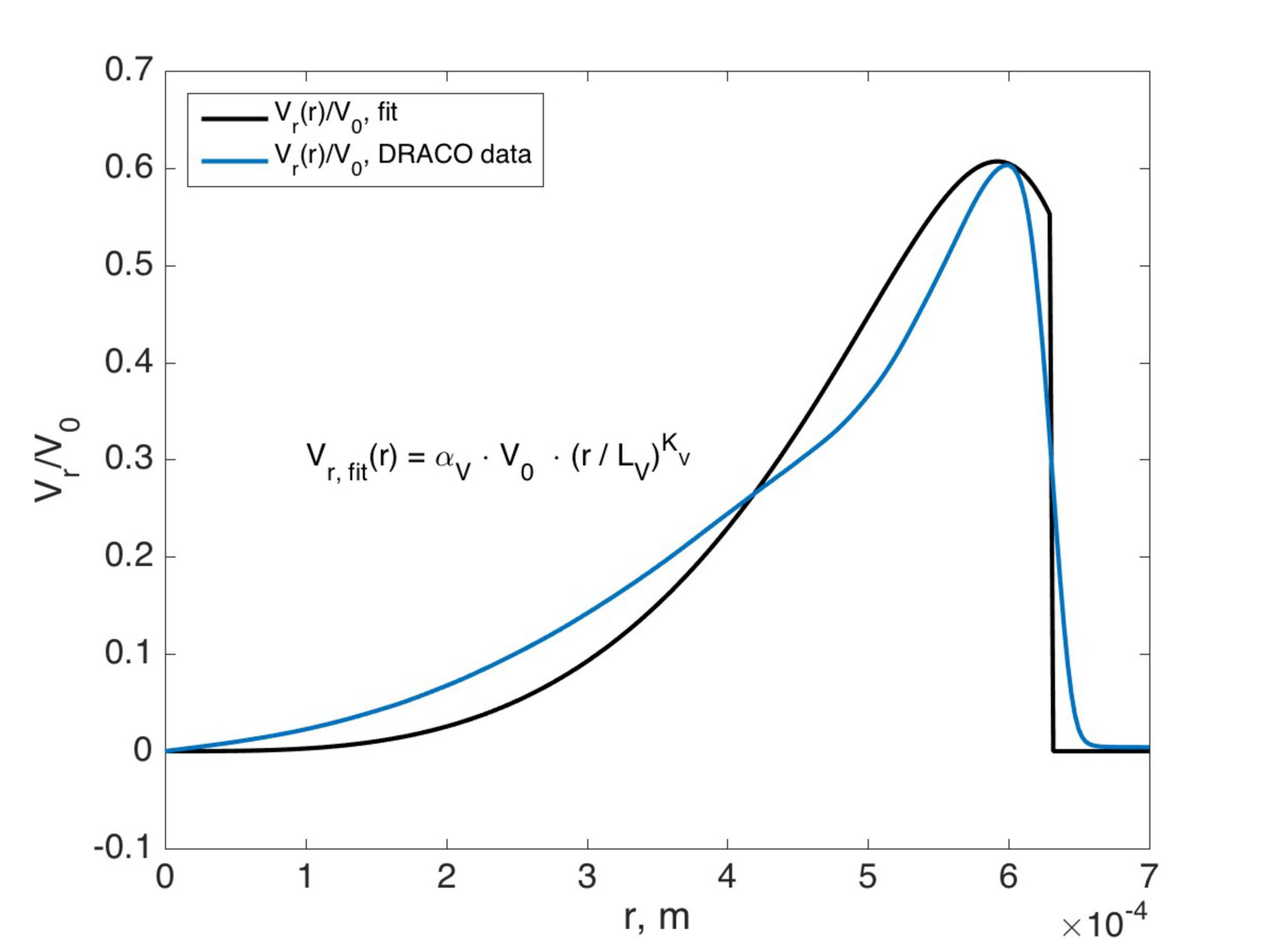}
\caption{Initial radial velocity profile for single plasma bubble from actual DRACO data (aqua) and resulting fit in form of trimmed power law (black).}
\label{fig:init2}
\end{figure}

These cuts provide density, temperature, flow, and magnetic field as a function of radius. They were fit to analytic functional forms for ease of initializing PSC, see Figures \ref{fig:init1} and \ref{fig:init2}.  The profiles consist of a uniform background (density $n_b$, temperature $T_b$) and multiple plasma bubbles with radial profiles described in detail by Table \ref{table:init}. When combining multiple bubbles and a background plasma, the velocities are averaged, weighted by the local density (to obtain a single flow speed at each location, even though multiple populations initially exist there).  Accordingly, a small correction factor $\alpha_V \approx 1.8$ is used to boost the speed to match the DRACO profiles. Two bubbles are separated by a distance $2L_z = 1.4 \, \rm mm$.

\begin{table}
\centering
\caption{Parameters for OMEGA-EP magnetized plume reconnection experiments}
\begin{tabular}{lr}
\tableline\tableline
Analytic fitting form: \\
\tsp $n$ $= n_0 \cdot \exp \left( -(r / L_N)^{K_N}\right)$\\
\tsp $T$ $= T_0 \cdot \exp \left( -(r / L_T)^{K_T} \right)$\\
\tsp $V = \alpha_V \cdot V_0  \cdot (r / L_V)^{K_V},  r < L_V $\\
\tsp $B = B_0 \cdot \exp \left( -((r-R_B)/L_B)^2 \right)$\\

\\

Fitting parameters: \\
\tsp $n_{0}$ (m$^{-3}$) & \sciexp{1.35}{26} \\
\tsp $L_N$ (m) & \sciexp{350}{-6} \\
\tsp $K_N$ & 2.8 \\
\tsp $T_{0}$ (eV) & 1050 \\
\tsp $L_T$ (m) &   \sciexp{640}{-6} \\
\tsp $K_T$ & 12 \\
\tsp $V_0$ (m/s) & \sciexp{6}{5} \\
\tsp $L_V$ (m) & \sciexp{630}{-6} \\
\tsp $K_V$ & 3.2 \\
\tsp $\alpha_V$ & 1.8 \\
\tsp $B_0$ (T) & 50 \\
\tsp $L_B$ (m) & \sciexp{43}{-6} \\
\tsp $R_B$ (m) & \sciexp{470}{-6} \\

\\
Dimensionless numbers: \\

\tsp $L_z / d_{i0}$	& 40 \\

\tsp $V_0 / C_s$ & 	2 \\

\tsp $B_0^2 / 8 \pi n_0  T_0$ &  0.05 \\

\tsp $\eta_0 = \nu_{ei0} / \omega_{ce0}$	& $\sim 0.05 $\\


\tsp $n_{b} / n_{0}$ & $\sim 0.005 $\\
\tsp $T_{b} / T_0 $ & 0.1 \\
\end{tabular}
\label{table:init}
\end{table}

We consider two geometry setups - two circular expanding plasma bubbles and two parallel sheets (magnetic fields in these sheets are antiparallel) moving towards each other. For two circular expanding plasma bubbles, $r = \sqrt{ (x-x_0)^2 + (z-z_0)^2}$, where x and z represent the distance from the bubble center perpendicular and along the direction of separation.  For two quasi-1D parallel colliding plasmas modelling a very long current sheet experiment at NIF, we choose $r = {\rm abs}(z-z_0)$, so that the initial condition is invariant along x.

These are translated into numerical parameters for the PIC code PSC. In the numerical implementation, all length scales ($L_T$, $L_N$, etc) are normalized to the half-separation $L_z$ to give a single universal profile, which is then controlled in comparison to the plasma kinetic scale through the single dimensionless parameter $L_z / d_{i0}$.  The magnitude of the B field is controlled by matching the plasma beta.  As is typical in PIC simulations, the electron-ion mass ratio and ratio of speed of light to electron thermal speed must be compressed.  This treatment matches the method of matching the ion-scale dimensionless parameters discussed in Ref. \cite{HEDP1}.

From the parameters above we obtain the following baseline dimensionless parameters, see Table \ref{table:init}.  These parameters are regarded as representative and we vary them in the simulation study.  In particular we vary $\eta_0$ to obtain various Lundquist numbers, and background density $n_b$ which controls the local ion skin depth where the plasmas collide.

As mentioned above, our PIC simulations are carried out using PSC \cite{GERMASCH}. We choose the parameters
$M_i/m_e = 100$, $T_{i0} = T_{e0} = 0.02 m_ec^2$, and
an initial 400 particles per cell. We choose the grid with 20 nodes per $d_{i0}$, where $d_{i0}=c/\omega_{\rm pi}$ is the ion skin depth, $\omega^2_{\rm pi} = 4 \pi e^2 n_0/M_i$ is the ion plasma frequency calculated for $n_0$. The total grid is $3200 \times 1600$ nodes. { Boundary conditions are periodic for both particles and fields.}
We perform a series of simulations with various values of collisionality parameter $\eta_0 =\nu_{ei0} / \omega_{ce0}$, where $\nu_{ei0}$ is the electron-ion collision frequency and $\omega_{ce0}= e B_0 / m_e c$ is the electron gyrofrequency, where $e$ is the electron charge, $m_e$ is the electron mass and $c$ is the speed of light. PSC implements a binary Coloumb collision operator, for further details please see \cite{GERMASCH}. Thus, we perform simulations for both geometries for $\eta_0 $ within a range from 0.0001 to 5.0. For the largest collisionality value the corresponding test simulation gives no larger than 15\% of error in terms of Spitzer current, according to $\rm Table \, 1$ in \cite{GERMASCH}. We also vary the background plasma density, $n_b$ in order to understand how the fluid effects may change the process of current sheet formation and reconnection. To our understanding, the background density has not been well-measured in the experiments we are trying to simulate. Within the computational model, it is best thought of as interesting proxy which controls how thin the current sheet can become in comparison to $d_i$. We choose the background density $n_b$ within a range from 0.001 to 0.04 times the maximum density in plasma bubbles, $n_0$.

\section{Simulation results}
\label{sec:results}

Let us first discuss the typical picture we observe in our simulations of magnetic reconnection. In the expanding bubble case, the approximate spatial location of the X-point is largely determined by the geometry of the bubbles. In the low-collisionality, low-background density case, after the collision of two magnetic ribbons,  we usually observe multiple X-points (typically, two or three of them) due to the excitation of the plasmoid instability. As a result, a few plasmoids are formed, and the reconnection completes {(Figure 3 shows this process at the midway of the reconnection, when the maximum reconnection rate occurs)}. In the parallel sheet case, there is no pre-imposed geometry defining an X-point, so the reconnection process must be triggered by the tearing instability. Eventually, in the low-collisionality, low-background density case, the two parallel ribbons undergo the plasmoid instability, and reconnect through multiple (from 6 to 11) X-points, and form a number of plasmoid-like structures, as demonstrated in Figure 4. 

\subsection{Tracked parameters}

We measure several parameters of the current sheet through the course of the simulation to quantify their role on reconnection. Typically these parameters are calculated in relation to ``local'' parameters evaluated near the current sheet, rather than the ``global'' parameters characteristic of the initial conditions.  

We calculate the Lundquist number using the following formula:

\begin{equation}
S =  \frac{L v_A}{\eta}= \frac{L}{d_i} \frac{1}{\eta_\ast},    
\end{equation}

\noindent where $L / d_i$ measures the length of the current sheet in local $d_i$ units.  $L$ is evaluated by dividing the total current sheet length ($L \sim 100\, d_{i0}$) by the number of X-points formed. We apply this definition of the current sheet length for both expanding bubble and parallel sheets geometries.  The dimensionless collisonality parameter, $\eta_\ast$, is calculated as follows:

\begin{equation}
\eta_\ast = \eta_0 \cdot \frac{n_{e \ast}}{n_{e0}} \cdot \left(\frac{T_{e0}}{T_{e \ast}}\right)^{\frac{3}{2}} \cdot \frac{B_0}{B_{\rm up}}.
\end{equation}

\noindent Here, again,
$\eta_0$ ranges from 
\, 0.0001 to 5.
$B_{\rm up}$ is the upstream magnetic field measured at time of maximum reconnection rate and is calculated as the maximum magnetic field in the square region of $20 d_{i0} \times 20 d_{i0}$ around the X point. The parameters specified with asterisks are taken directly from the code at the X point. We do not have a constant Lundquist number $S$ at the X point in the reconnection process due to plasma compression and drop of the temperature due to transfer of colder plasma to the X-point. Furthermore, we see a sufficient drop in the Lundquist number (by as large as a factor of ten) at later stages of the current sheet evolution. For purposes of comparing multiple simulations, we characterize them by the Lundquist number at the X point (in case of multiple X points we take the one with the highest physical reconneсtion rate) at the time of the maximum reconnection rate.

We calculate the normalized reconnection rate using the following formula:

\begin{equation}
\frac{{\rm d} \psi}{\rm dt} = \frac{E_{y \ast}}{v_{\rm up} B_{\rm up}},    
\end{equation}

\noindent where $v_{\rm up}=B_{\rm up}/B_0 \cdot \sqrt{n_0/n_\ast} \cdot v_{\rm A0}$ and  $B_{\rm up}$ are the Alfven velocity and magnetic field calculated from the upstream magnetic field value. $E_{y \ast}$ is the value of $y$-component of the electric field in the X-point, taken directly from the code. Note that we take into account the effect of the magnetic flux pileup \cite{Karimabadi2011}. The importance of such an effect will be discussed later.

We measure the width $\delta$ of the current sheet by the following procedure: 1) find the X-point; 2) find the location of two minima of $J_y$ on either side of the X-point along the 1D cut at $x=x_\ast$ in Z direction; 3) take the half-width between these two points. Typically we compare $\delta$ to the local $d_i$, evaluated in the X-point as $d_{i \ast}/d_{i0}=\sqrt{{n_{e0}}/{n_{e \ast}}}$. In order to calculate the Sweet-Parker width $\delta_{\rm SP} = L/S^{1/2}$, we use the usual formula $\delta_{\rm SP}/d_{i0}={L}/{(d_{i0} S^{1/2})}$, and $S$ is evaluated as above. 

In order to track the evolution of the magnetic flux in the case of colliding bubbles, we calculate the following integral:

\begin{equation}
    \psi(t) = \int_{z_\ast(t)}^{z_{\rm bound}} B_x (x_\ast(t),z,t) dz,
\end{equation}
\noindent where $B_x$ is the inflow ($x$) component of the magnetic field, $x_\ast$ and $z_\ast$ are coordinates of the x-point (they may not be the same for various moments of time), and $z_{\rm bound}$ is the boundary coordinate of the simulation box. This allows us to define time needed for the reconnection of 80 \% of the initial flux (denoted $t_{\rm 80}$) from the equation $\psi(t_{\rm 80})=0.2\, \psi(0)$. 

\subsection{The role of resistivity}

To start with, let us analyze the series of PIC simulations with fixed background density ($n_b/n_0 = 0.005$) and varying resistivity.

Figure \ref{fig:summarybubble} presents a summary of the various reconnection behavior in the expanding bubble case. For collisional reconnection ($S<10^3$), single X-point reconnection is observed, and the time to reconnect 80\% of the flux, $t_{80}$, is three times higher than in the collisionless ($S\gg 10^3$) case. In the collisionless reconnection case, we observe a complete reconnection through multiple X points due to the plasmoid instability. The physical behavior dramatically changes for the Lundquist numbers around $S \approx 10^3$. For instance, there is a minimum in the reconnection rate curve (red crosses) for this $S$ value, which separates the diffusive reconnection regime ($S < 5\cdot10^2$) and collisionless reconnection regime $S > 3 \cdot 10^3 $. This transition is captured by both Sweet-Parker theory (blue circles) and the simulation current sheet width (blue line; we will refer to this as to $\delta/d_i < 1$ criteria). 
The lower subplot shows the time of the 80\% of the initial flux being reconnected, which also clearly illustrates the transition between collisionless ($S>3 \cdot 10^3$), collisional ($S\approx 5\cdot10^2 - 3 \cdot 10^3$), and diffusive reconnection ($S<5\cdot 10^2$). The transition between plasmoid and single X-point reconnection happens around $S\approx 10^3$. At and below such values of $S$, we expect that the Sweet-Parker current sheet is realizable, as predicted by recent theories of evolving current sheets \cite{Pucci2014,Comisso2017,Huang2017}.

Figure \ref{fig:summaryparallel}, which corresponds to the case of parallel sheets, demonstrates that the overall picture is somewhat similar to the expanding bubble case. The transition in terms of the maximum reconnection rate occurs around $S\approx 10^3$ - smaller values lead to the diffusive reconnection regime, larger values to collisionless reconnection. 
Sweet-Parker theory suggests the transition to sub-ion skin depth scale ($\delta_{\rm SP}/d_{i0}<1$) happens around $S \approx 7\cdot10^2$, or, from the $\delta/d_i < 1$ criteria, $S\approx 3\cdot10^3$. Again, the transition between plasmoid and single X-point reconnection occurs around $S\approx 10^3$. On the other hand, in contrast to the bubble case, we see that the reconnection is stalled for $S<2\cdot10^3$; the 80 \% of the flux never reconnects through the simulation time. What happens instead is a hydrodynamical rebound of the two plasmas which completely shuts off the plasma inflow to the reconnection layer, a process illustrated in Fig. \ref{fig:collisional_rebound}. We track the evolution of $B_x$, $v_{\rm ez}$, and $n_e$ in the vicinity of the X point. Figure \ref{fig:collisional_rebound}a presents the initial condition with two magnetized ribbons moving towards each other. Fig. \ref{fig:collisional_rebound}b demonstrates the moment of the maximum magnetic flux pile-up, which is usually observed right before the beginning of the reconnection process. Figure \ref{fig:collisional_rebound}c shows that some part of the initial flux is reconnected, but we also see that the inflow velocity has gone down to zero. This means that the magnetic field lines stop flowing toward each other, after which these two ribbons rebound and begin moving apart (Fig. \ref{fig:collisional_rebound}d). {After the rebound, the only source of the flux dissipation is small resistive diffusion (in all but the diffusive reconnection regime, where it may be significant source of the magnetic flux decay).}

Reconnection rates observed in our simulations are in a range $0.1 - 0.5\, V_{A \ast} B_\ast$, where the values with asterisk are the upstream Alfv\'en velocity and magnetic field magnitude. The contribution of flux pileup may vary from a factor of one in strongly collisonal case to a factor of ten in the collisionless case. The parallel geometry case in general gives slightly larger flux pileup and, subsequently, slightly smaller normalized rates. Thus, collisional effects are important for the process of the current sheet formation, preceding the reconnection process itself.     

Throughout our simulations, we see that there are both quantitive and qualitative differences between the two geometry setups, the most distinctive of which is the effect of the rebound of the magnetized ribbons, which is present more clearly in the parallel sheet case.

\begin{figure}
    \includegraphics[width=0.99\linewidth]{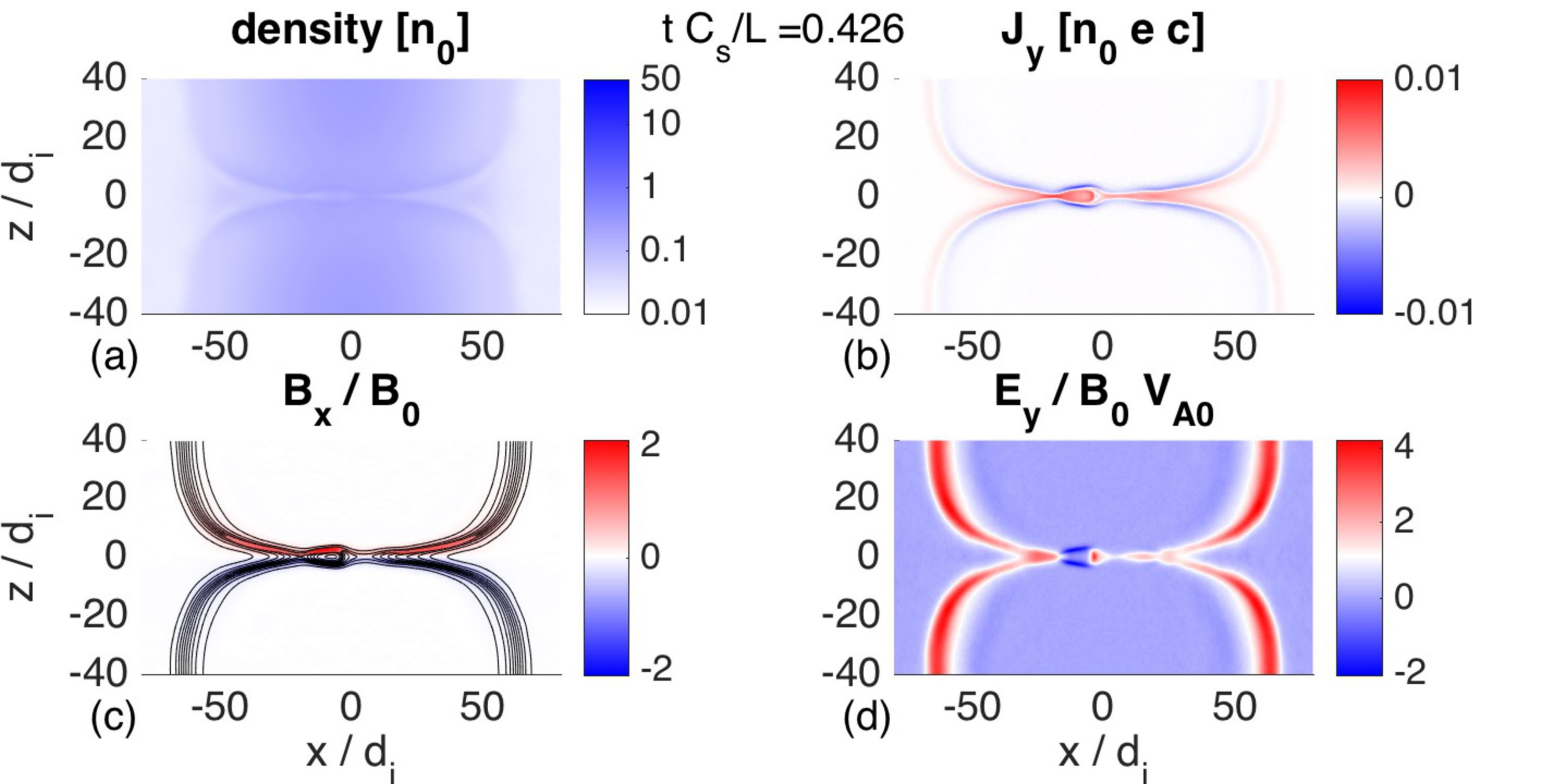}
\caption{Bubble collision case for $S\sim 8 \cdot 10^3$. The plasmoid structure in the center of the collision is seen, with two X points around it. Heated (three times the initial temperature) and dense (as dense as the initial maximum value in plasma bubbles) plasma is observed around the plasmoid structure. Maximum magnetic field is no less than two times larger than the initial one due to the flux pileup.}
\label{fig:summarybubble}
\end{figure}

\begin{figure}
    \includegraphics[width=0.99\linewidth]{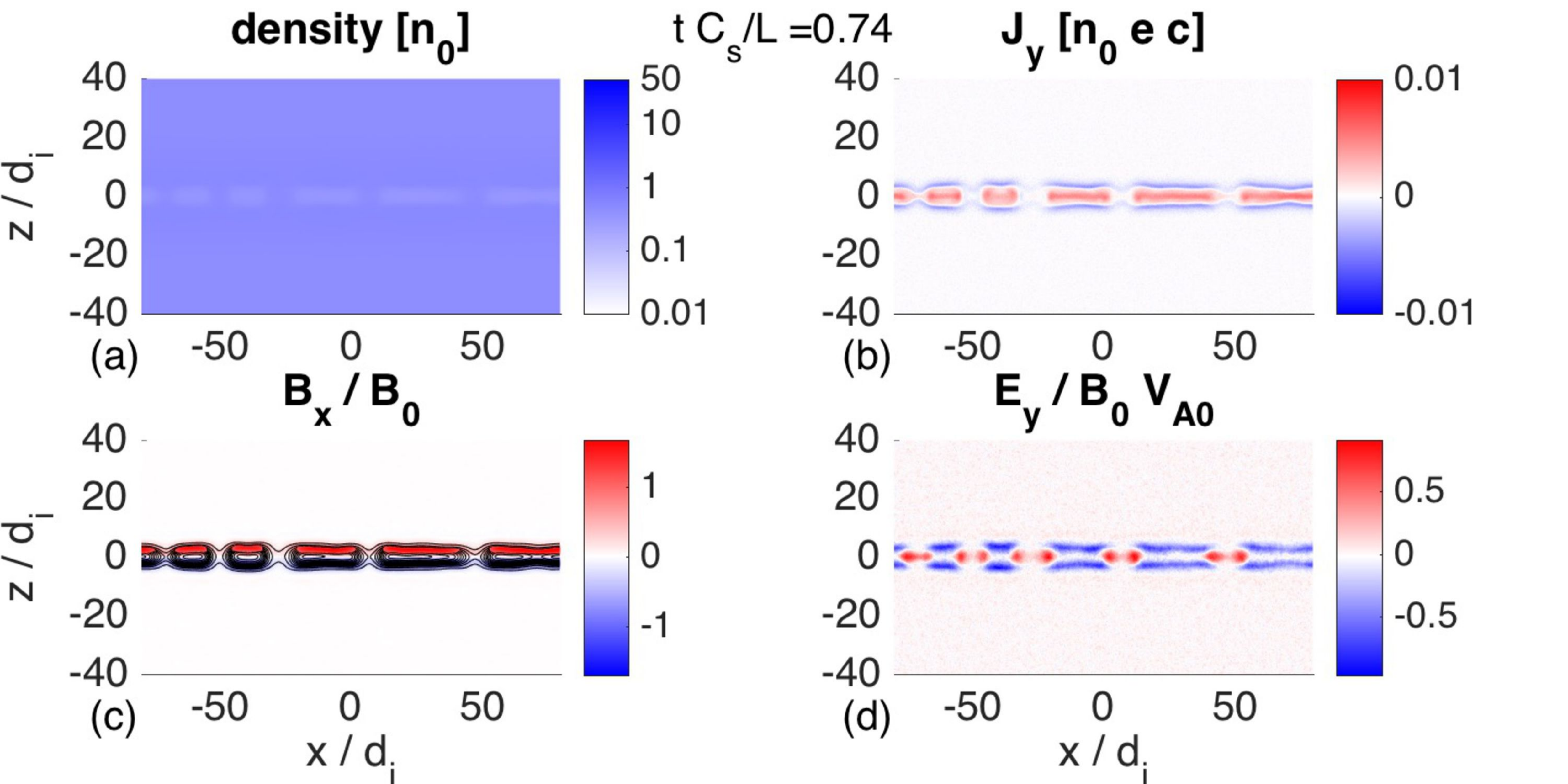}
\caption{Parallel collision case for $S \sim 1.1 \cdot 10^4$. A large number of plasmoid structures (and surrounding X points) is observed. Magnetic flux pileup and plasma heating in the vicinity of the plasmoid structure are again seen. }
\label{fig:summaryparallel}
\end{figure}

\begin{figure}
    \includegraphics[width=0.99\linewidth]{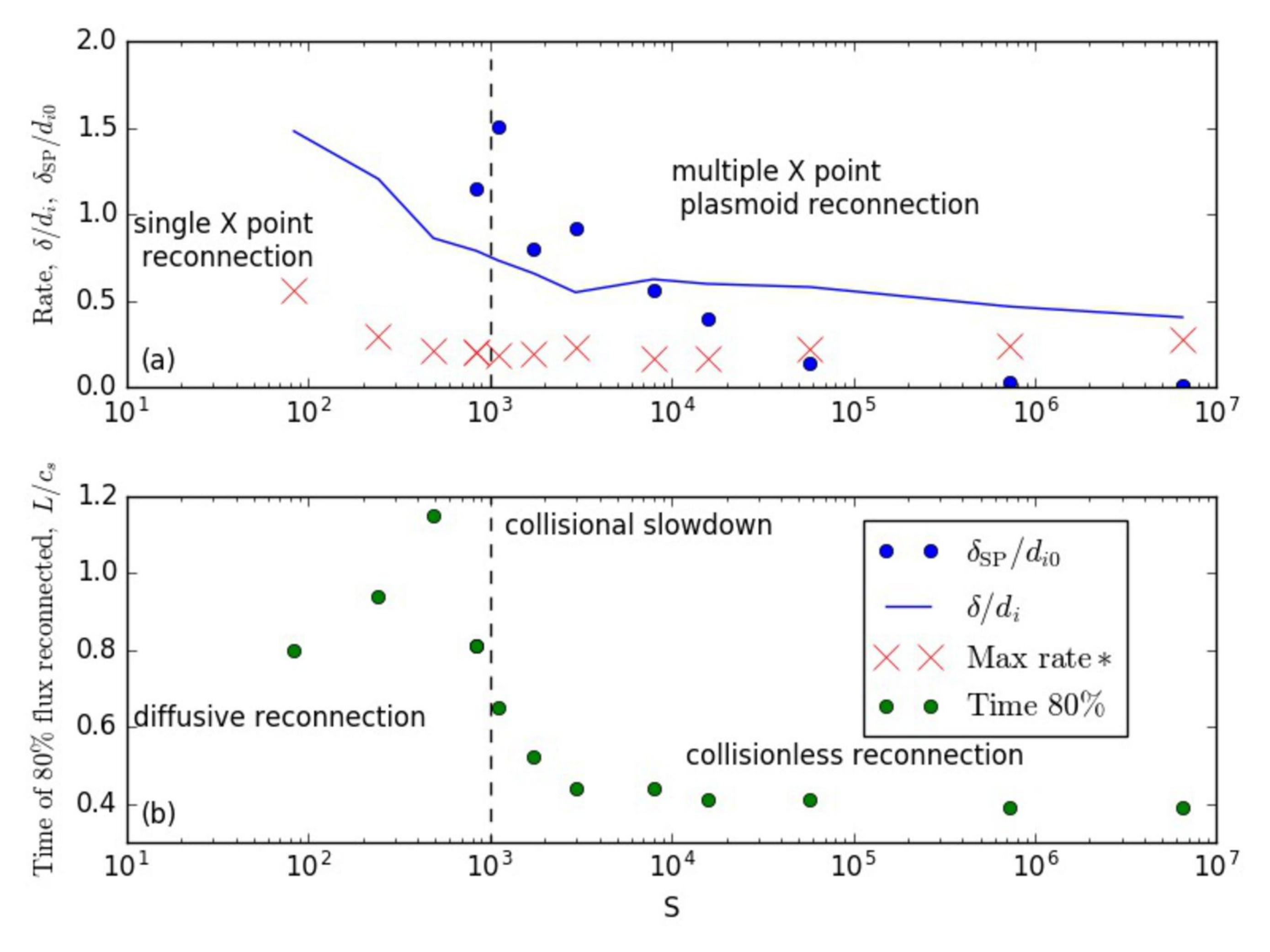}
\caption{Summary for the expanding bubble case.
(a) The transition from a single X point reconnection to multiple X point plasmoid reconnection. Red crosses stand for the reconnection rate, blue line shows the $\delta/d_i$ value for different S simulations, blue points represent the $\delta_{\rm SP}/d_{i0}$ values. The transition from slow reconnection to fast is captured by SP and $\delta/d_i$.
(b) The transition from diffusive reconnection ($S<5\cdot 10^2$) to collisional ($S\approx 5\cdot10^2 - 3 \cdot 10^3$) and collisionless ($S>3 \cdot 10^3$) reconnection in terms of the time needed for 80\% of the initial flux to reconnect (green points).}
\label{fig:summarybubble}
\end{figure}

\begin{figure}
    \includegraphics[width=0.99\linewidth]{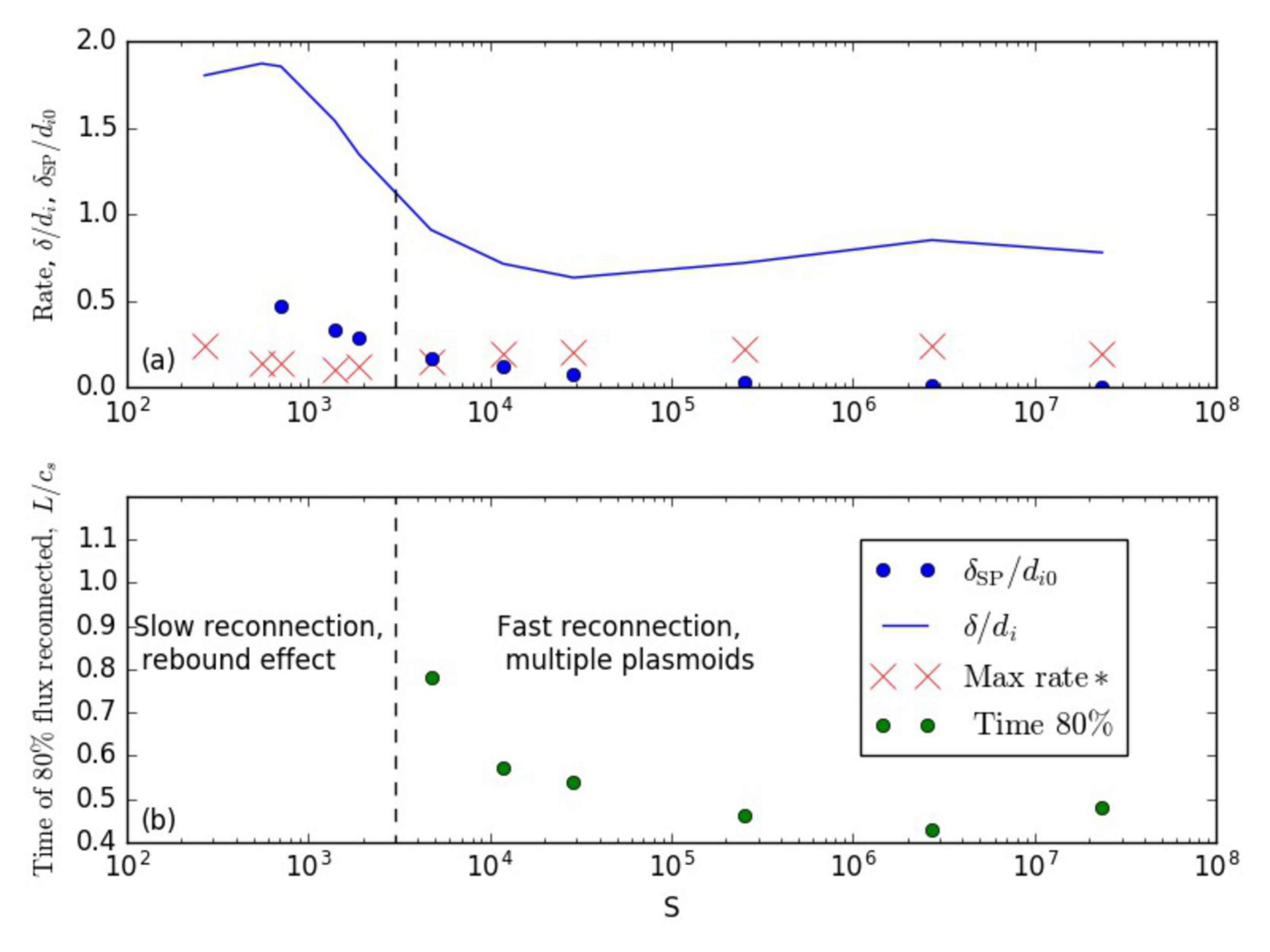}
\caption{Summary for the parallel sheet case for various collisionality values. 
(a) The transition from a single X point reconnection to multiple X point plasmoid reconnection. Red crosses stand for the normalized reconnection rate, blue line shows the $\delta/d_i$ value for different $S$ simulations, blue points represent the $\delta_{\rm SP}/d_{i0}$ values. (b) The transition from slow reconnection due to the rebound to collisional reconnection ($S\approx 2\cdot10^3 -   10^4$) and collisionless ($S> \cdot 10^4$) reconnection in terms of the time needed for 80\% of the initial flux to reconnect (green points). }
\label{fig:summaryparallel}
\end{figure}

\begin{figure}
    \includegraphics[width=0.49\linewidth]{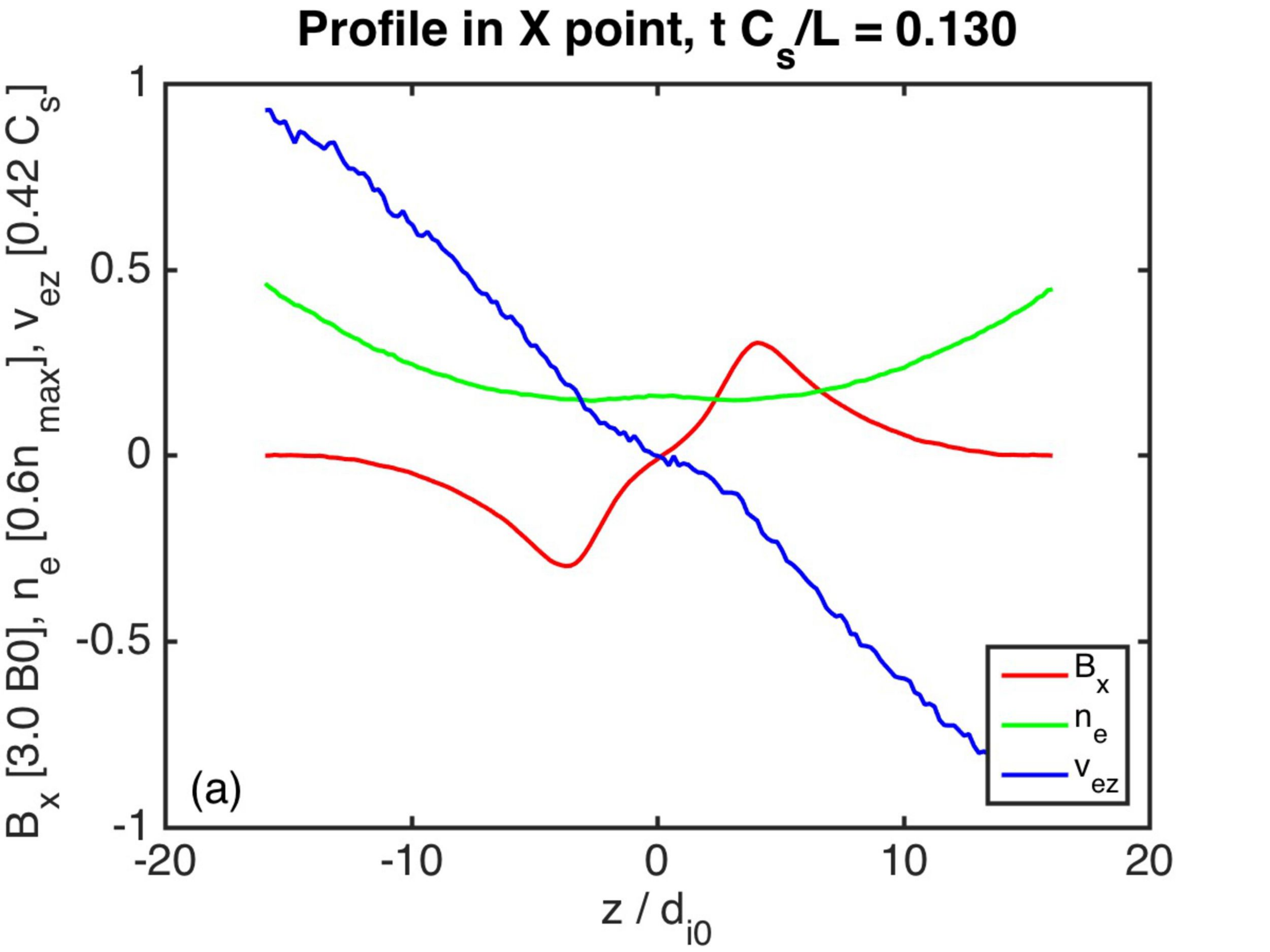}
    \includegraphics[width=0.49\linewidth]{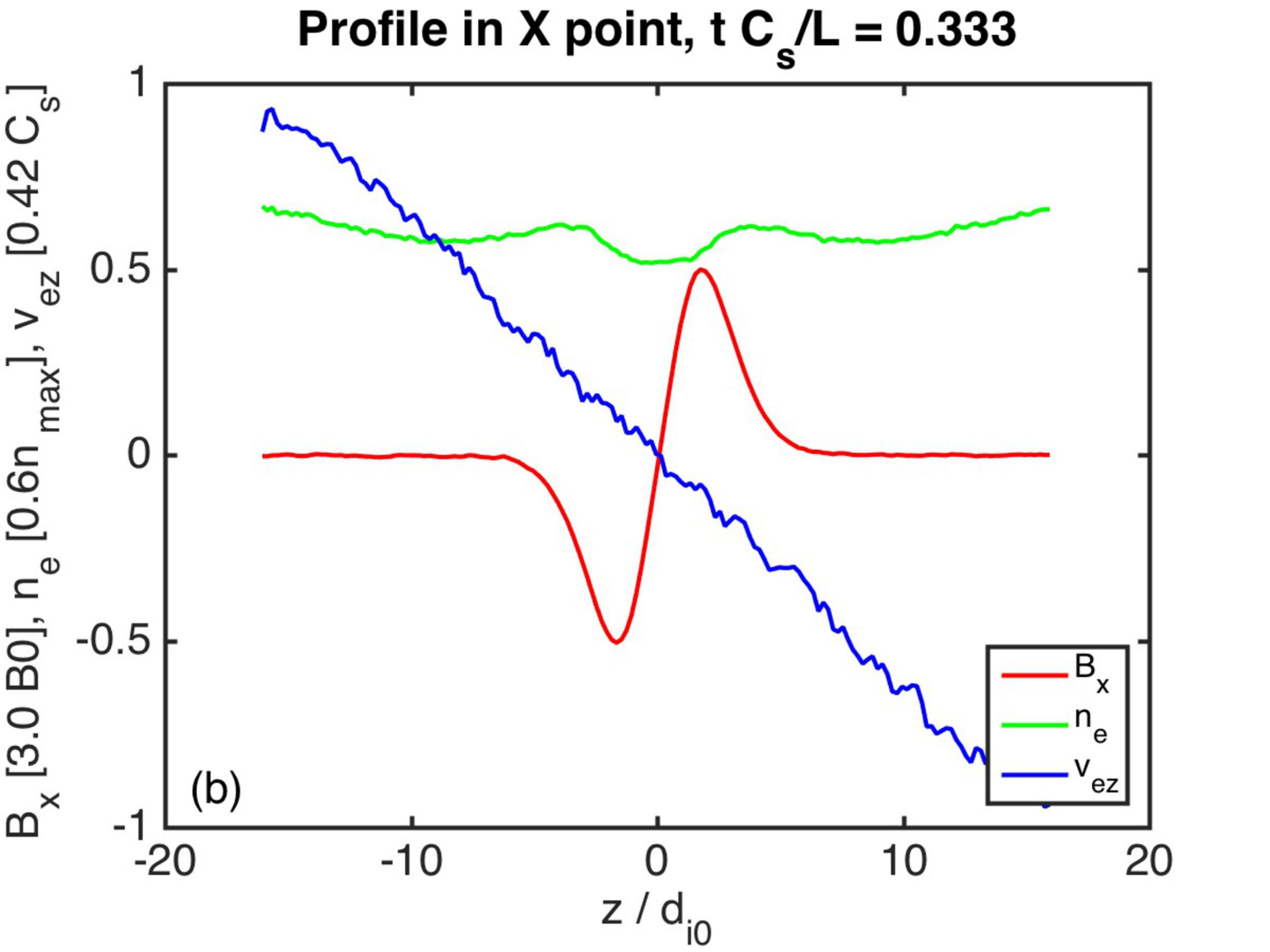}\par
    \includegraphics[width=0.49\linewidth]{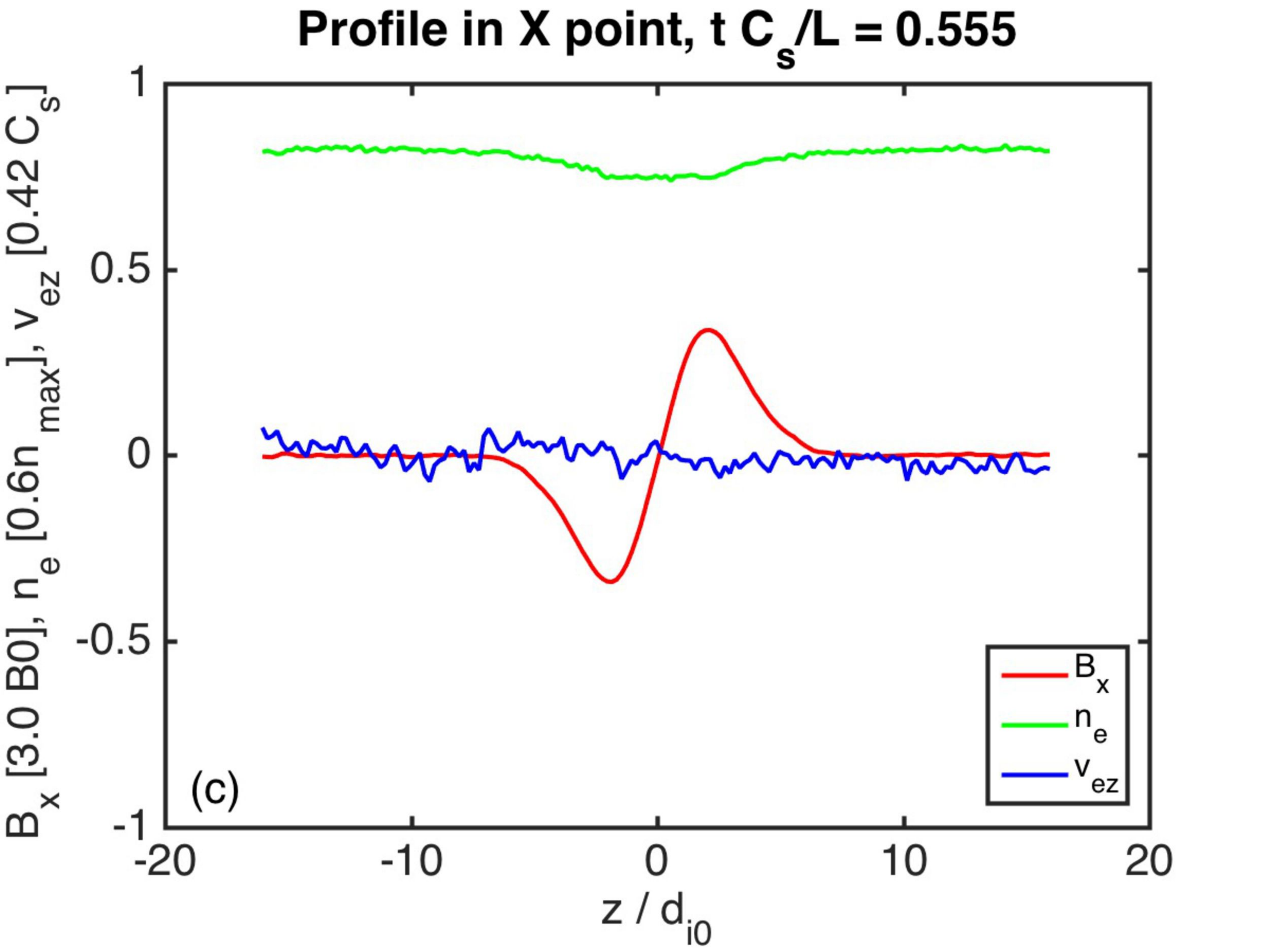}
    \includegraphics[width=0.49\linewidth]{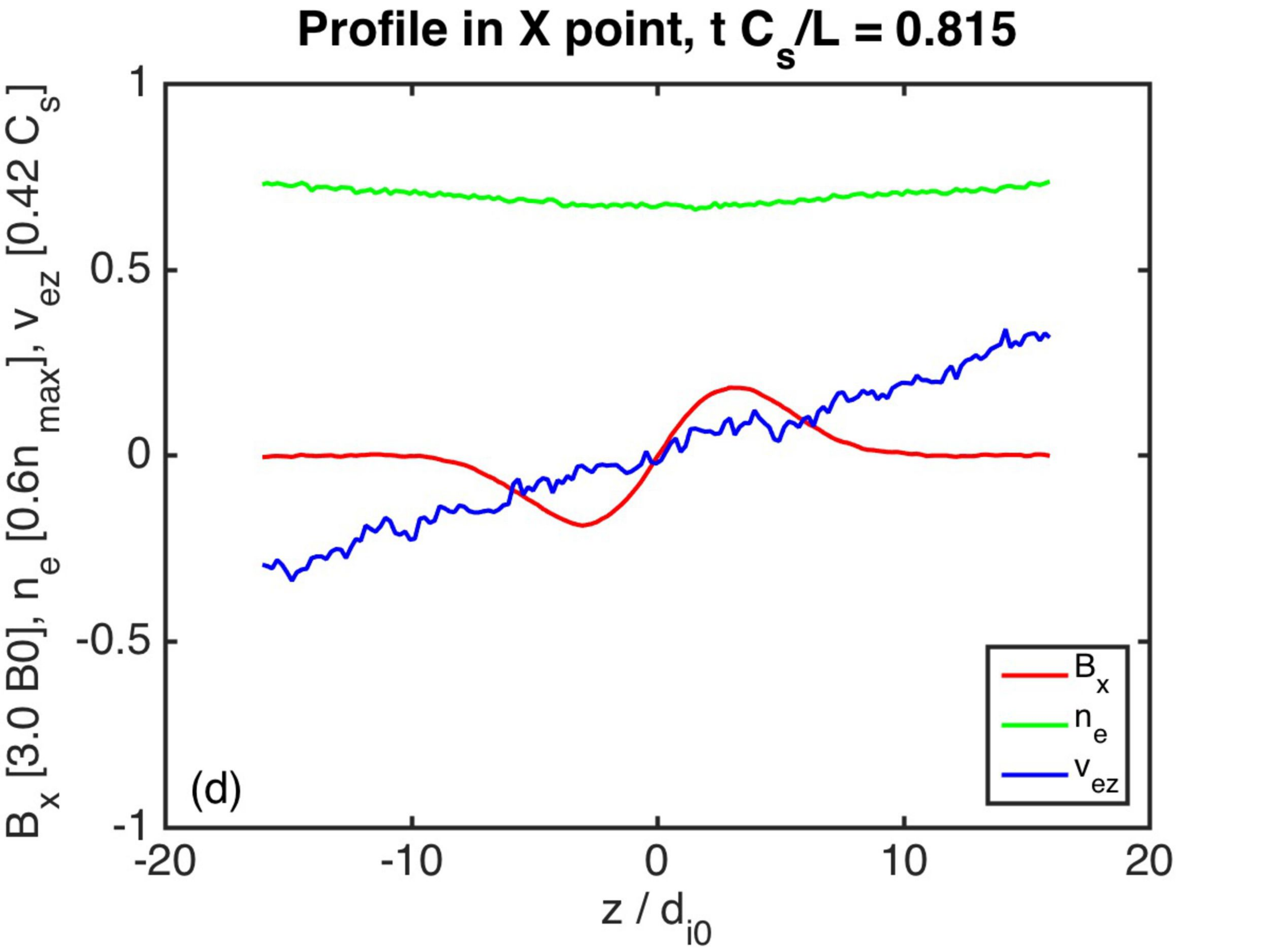}
\caption{Evolution of $B_x$, $v_{\rm ez}$, and $n_e$ in the X-point for the parallel collision simulation with $S\approx 10^3$, illustrating the shutdown by hydrodynamic rebound of two two plasmas in a collisional regime. (a) $t=0.13 L/C_s$ - near the initial moment of the simulation; (b) $t=0.33 L/C_s$ - maximum flux pileup time; (c) $t=0.56 L/C_s$ - the time at which the plasma inflow towards the X-point completely stops, and (d) $t=0.81 L/C_s$ - the rebound effect is observed, where reversal of the inflow is seen, along with the resistive diffusion of two magnetized ribbons. The amount of flux reconnected/diffused out is no more than 40 \%.}
\label{fig:collisional_rebound}
\end{figure}

\begin{figure}
    \includegraphics[width=0.49\linewidth]{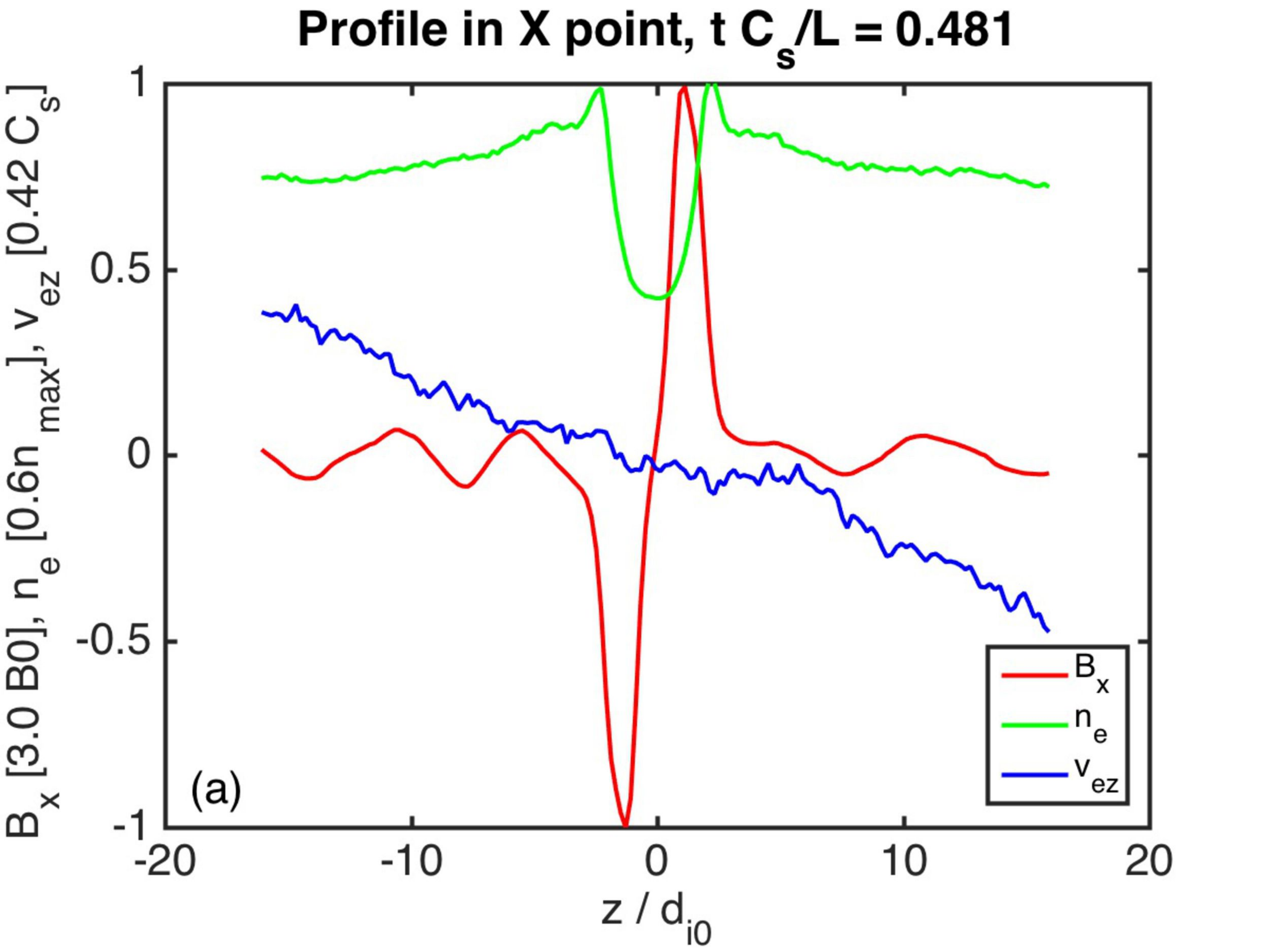}
    \includegraphics[width=0.49\linewidth]{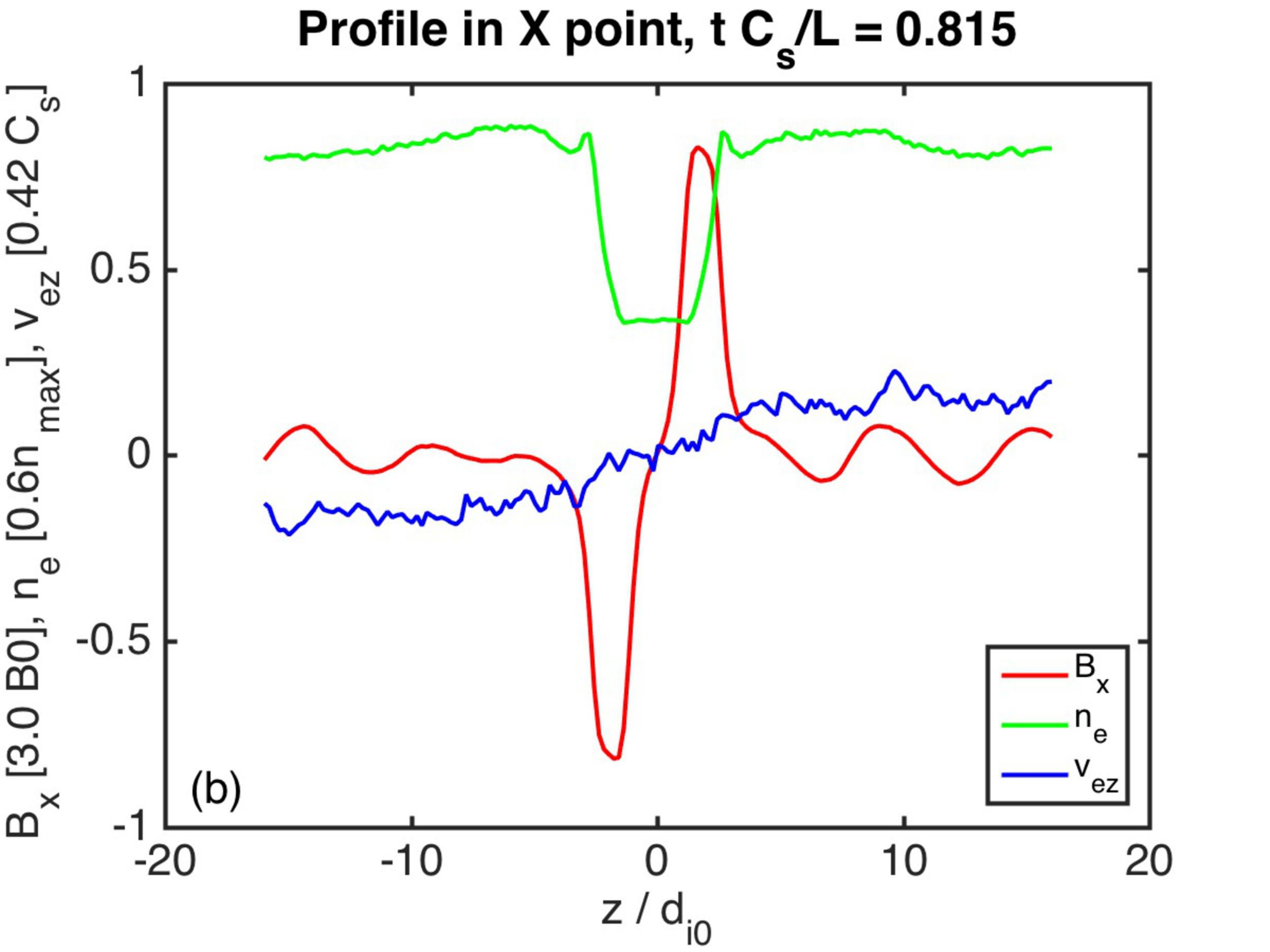}\par
    \includegraphics[width=0.99\linewidth]{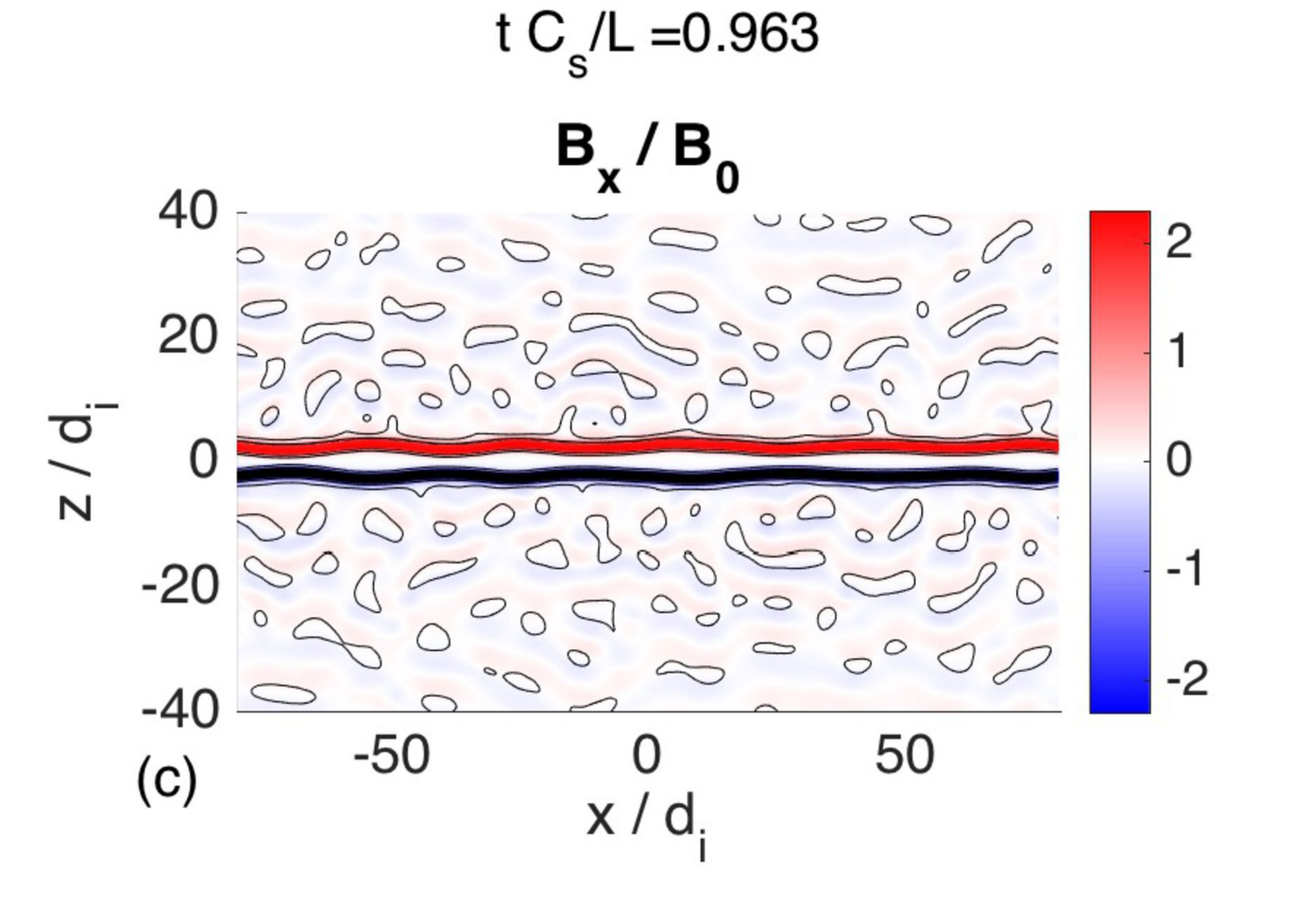}    
\caption{Evolution of $B_x$, $v_{\rm ez}$, and $n_e$ in the X-point for the parallel collision simulation with $S=\infty$ and $\rm n_b/n_0=0.01$ for (a) $t=0.48 L/C_s$ - maximum flux pileup time and (b) $t=0.81 L/C_s$ - the rebound effect is observed, where the outflow from the X-point is seen. Subplot (c) shows the final stage of the simulation, where the reconnection is stalled. The amount of flux reconnected in this simulation is no more than 10 \%.}
\label{fig:collisionless_rebound}
\end{figure}

\subsection{The role of the background density}

Above, we have demonstrated the important role fluid effects play in the considered magnetic reconnection setup. In order to further investigate the influence of fluid effects on the formation of the current sheet and subsequent magnetic reconnection, we consider another parameter which may be important from the fluid standpoint - the background plasma density.

 For the expanding bubble geometry, a reasonable amount of background density ($< 5\%$) does not seem to influence the reconnection process -- the additional background density just delays the time of the maximum reconnection, and all other physical parameters of the reconnection process appear to be the same. For a high background density (for instance, at 10\%), the current sheet thining process stops at $\delta/d_{i} \approx 3$, which completely stalls the onset of fast reconnection even in collisionless simulations. 

\begin{figure}
\includegraphics[width=0.99\linewidth]{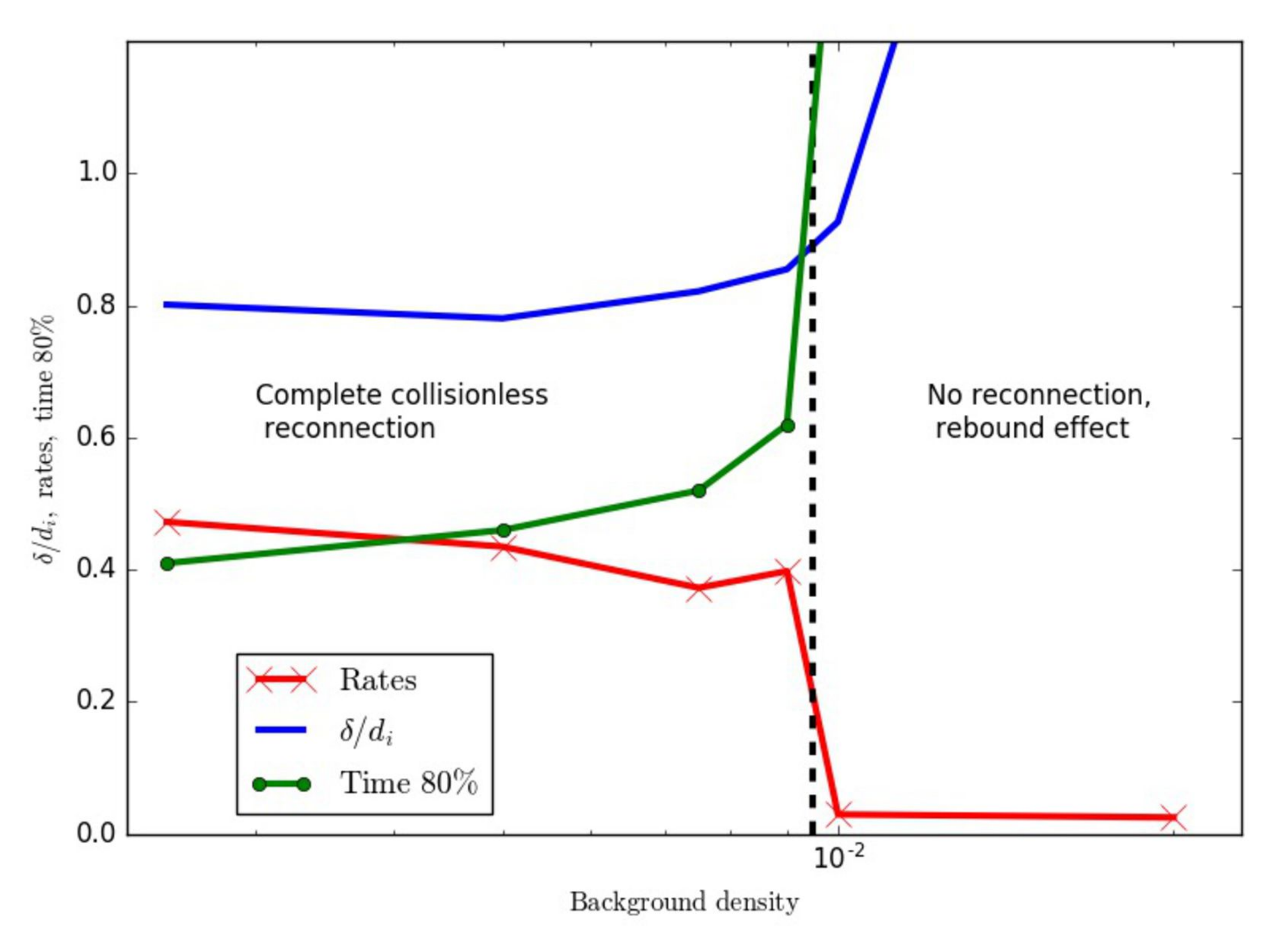}
\caption{Summary figure for the collisionless parallel collision case for various background density values. Green circled line shows the time needed for the 80\% of the initial flux to reconnect. Blue line depicts the $\delta/d_i$ values for various background densities.}
\label{fig:summarynnb}
\end{figure}

However, in the parallel sheet case, the additional background density may completely stop the reconnection process due to the rebound effect, similar to the rebound effect in the collisional case described in previous section, as was demonstrated in Figure \ref{fig:collisional_rebound}. Figure \ref{fig:collisionless_rebound} represents the process of rebound of two magnetized plasma ribbons in collisionless case. At the time of the maximum flux pileup the reconnection process starts, but then it is rapidly stopped by the reversal of the plasma inflow to the X-points. Figure \ref{fig:summarynnb} shows how the reconnection process is influenced by the background density in the parallel sheet case. The $\delta / d_i <1$ criteria for the fast reconnection onset again reproduces the value of the background density where the transition between fast and slow reconnection (or even no reconnection) happens -- around $1\%$ of the nominal density. The Sweet-Parker theory is indeed unable to catch such transition, as it occurs independent of collisionality.

When we introduce some finite collisionality into the simulation ($S \sim 3 \cdot 10^5$), there is a regime when {\it both} collisionality and additional background density may lead to full reconnection, even though, as we showed before, large collisionality or large background density by themselves substantially reduce the reconnection (see Figure 11).

\subsection{Tearing mode growth rates}

As we have emphasized before, geometry plays a vital role in triggering of the reconnection process. In the expanding bubble case, the X-point geometry is already there, so the reconnection occurs almost regardless of the plasma fluid effects - unless we put a considerable amount of background plasma between two bubbles. However, in the parallel sheet case, there is no imposed X-point geometry, so the X-point has yet to be formed. The process which triggers the formation of the X-point or multiple ones is usually associated with the tearing instability. In the parallel sheet case, it is possible to track the growth of such an instability, taking a Fast Fourier Transform of $B_z(x,z)$ along the x axis. The harmonics of such a decomposition grow in time, and saturate at some point. Growth rates are extracted from the simulations by fitting the mode growth rate to an exponential. Following the theoretical plasmoid instability description \cite{ComissoPoP2016}, we expect that Fourier harmonic with the largest linear growth rate will
correspond to the number of islands we observe in the magnetic field
map after the reconnection process. Though the linear growth rate works only for short times, it allows us to identify the fastest growing harmonic and it usually has the same number as the number of magnetic islands we observe during the time of peak reconnection rates. The final number of plasmoid-like structures may be smaller due to the coalescence of some of the neighbouring plasmoids.

The results for the observed growth rates in the parallel sheet case are the following. For Lundquist numbers $S>10^3$ the linear growth rates saturate at $\gamma L/C_s \approx 30$, remaining unchanged all the way to the completely collisionless regime ($S = \infty$). Going into the collisional regime, we find that the collisionality may reduce the growth rates by a factor of 4 by $S \sim 300$. The background density appears to be unimportant for the linear growth rate values, though there is a slight decrease for $n_b/n_0=0.02$, where $\gamma L/C_s \approx 20$.

Let us compare our results to the analytical estimates of the growth rates evaluated from the resistive tearing theory \cite{Furth1963} - $\gamma_{\rm max} L/C_s \approx (0.5 L_{\rm z}/\delta) (V_A /C_s) (0.5 L_z/L \cdot S)^{-1/2}$ which is no more than 0.5 for all S values considered. This indicates that reconnection is \textit{not} driven in HED systems by a resistive tearing instability. However, Hall MHD results are known to increase growth rates, and indeed using the results from Ref. \cite{Baalrud2011}, we can estimate growth rates: $\gamma_{\rm Hall MHD} L/C_s \sim 10^1$ range for $S\approx 10^3$, which is in much better agreement with simulations. However, for the highest Lundquist numbers ($S > 10^4$), $\gamma_{\rm Hall MHD} L/C_s$ overestimates the growth rate at least by two orders of magnitude. The collisionless tearing rate $\gamma L/C_s \approx \sqrt{\pi}/2^{3/2}\cdot (C_s/\delta \Omega_e)^{5/2} (\Omega_e L /C_s) (T_e+T_i)/T_e$, as defined by Ref. \cite{Coppi1966}, also gives growth rates around the one observed in simulations, but peaks at even larger values (up to $\approx 2 \cdot 10^2$) for large $S$ simulations. Overall, we may conclude that the resistive tearing is too slow to be present in our simulations which were conducted for $\sim L/C_s$, which is the ballistic time of the bubble expansion. However, values of collisionless tearing seem to be at least an order of magnitude larger than ideal tearing, and the approximate criteria $\gamma L/C_s > 10$ may be applicable in order to understand whether we observe tearing and plasmoid formation in our simulations. The importance of two-fluid effects is indicated by the approximate agreement with Hall MHD growth rate estimates. Our simulations also show that the Hall term has an important contribution (up to 80\% of the resulting electric field for $S>10^4$) to the generalized Ohm's law in the current sheet, in agreement with the simulations in \cite{HEDP1}.

\subsection{Interpretation of Rosenberg et al. (2015)}

\begin{figure}
\includegraphics[width=0.99\linewidth]{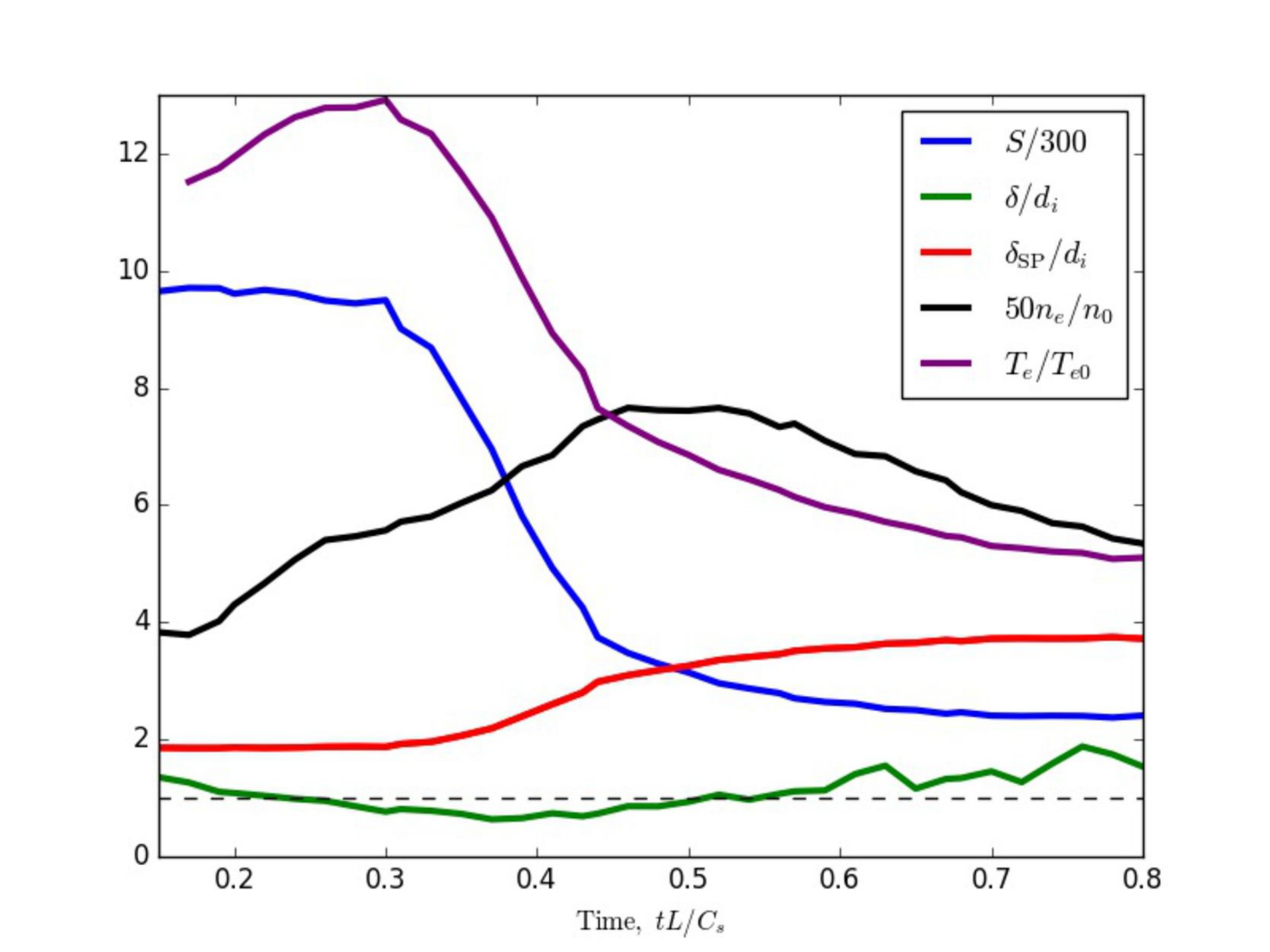}
    \par
    \includegraphics[width=0.99\linewidth]{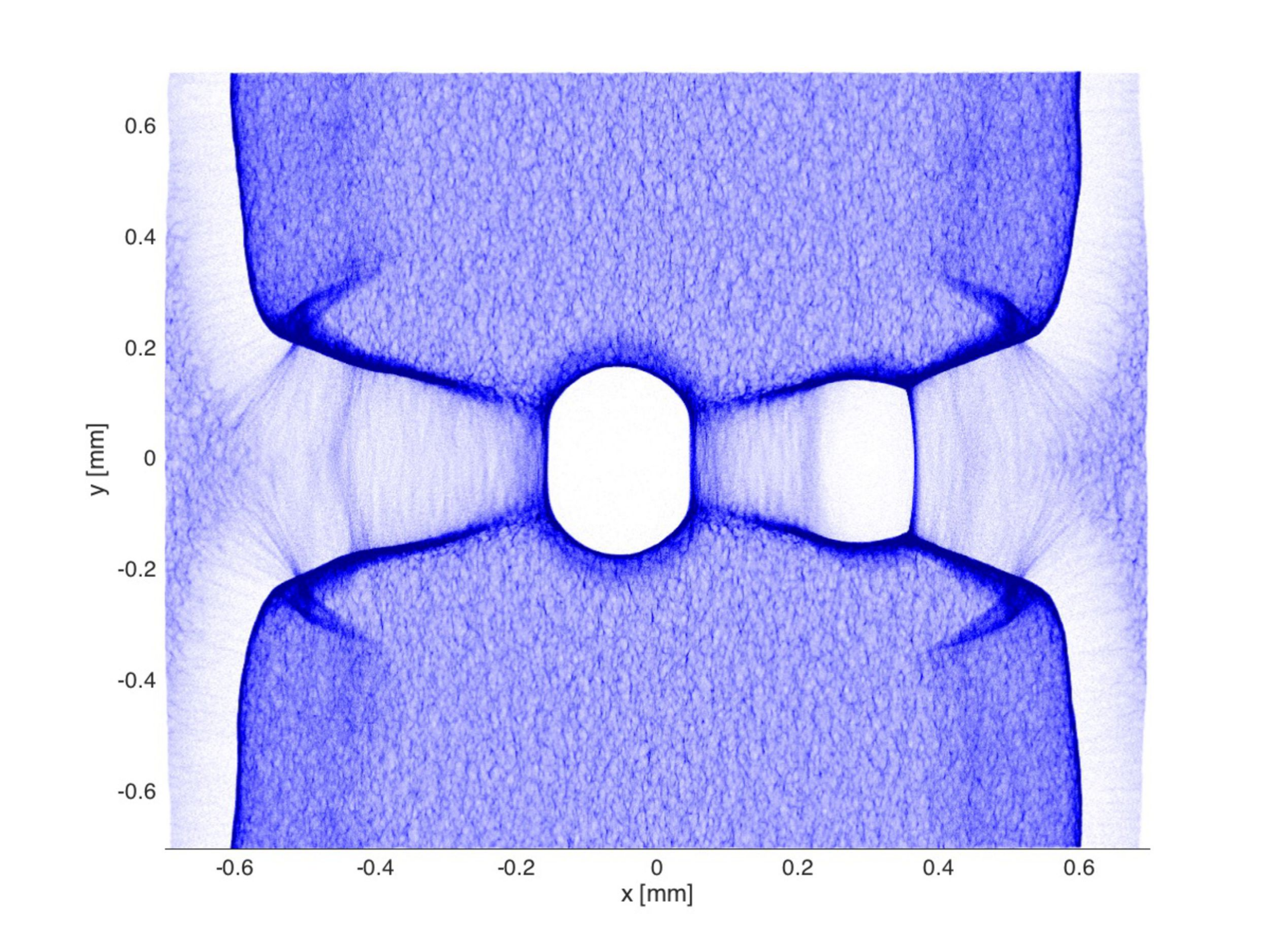}
\caption{(a) Evolution of Lundquist number, current sheet width, Sweet-Parker width, electron density, and electron temperature with time for the simulations that mimic parameters of the experiment \cite{Rosenberg2015}. Simulation with $\eta_0 = 0.3$ reproduces Figure 3 from Rosenberg et al. (2015) and appears to be in agreement with it. (b) Proton radiography reconstructed from simulation with $\eta_0=0.3$ for $t=1 {\rm ns}$. Plasmoid structures of around 400 microns in diameter are seen, which are not present at proton radiography by Rosenberg et al. (2015).}
\label{fig:rosenberg2015}
\end{figure}

Our results are relevant for the interpretation of recent experiments. For instance, in Ref. \cite{Rosenberg2015}, the experimemtal evidence for the possibility of a sudden slowdown of the reconnection was discussed. Their model implies that the increase in collisionality was the reason for the incomplete reconnection. We attempt to reproduce these results with the present PIC simulation model. First, as mentioned before, we initiate our PIC simulations using the bubble profiles from DRACO simulations in Ref. \cite{RosenbergPoP2015}. We track the evolution of Lundquist number at the X-points as well as other parameters (current sheet width $\delta/d_i$, Sweet-Parker width $\delta_{\rm SP}/d_i$, electron density, and electron temperature). We believe that the simulation with $\eta_0=0.3$ and bubble geometry reproduces the experimental setup used in Ref. \cite{Rosenberg2015}, as it reproduces the evolution of $S(t)$ in the X point as well as other current sheet parameters from the radiation hydrodynamics simulations in Ref. \cite{RosenbergPoP2015}. We observe the decline of $S$ up to a factor of three, see Figure \ref{fig:rosenberg2015}a, in agreement with \cite{Rosenberg2015}. In both Ref. \cite{Rosenberg2015} and our study temperature drops from the peak value by the factor of two during 0.5 ns. Electron density growth is present in both studies, being a factor of two in our study and factor of five in Ref. \cite{Rosenberg2015}, though they lie within the uncertainty range of each other. We performed the sensitivity test and checked that these results are almost independent from the collisionality parameter $\eta_0$ - simulations with $\eta_0=0.2$ and $0.4$ also show fast plasmoid reconnection with similar current sheet parameters. Our definition of Lundquist number includes the flux pileup effect, which was not the case for Ref. \cite{Rosenberg2015}, thus we overestimate our Lundquist numbers typically by a factor of two - all simulations with smaller $\eta_0$ will indeed show fast plasmoid reconnection.
\, However, in Ref. \cite{Rosenberg2015}, plasmoids were not observed. In our simulations, however, we see that plasmoids are formed and the proton radiography pictures reconstructed from the simulations show these plasmoid structures as clear circular zero-proton-fluence regions, see Figure \ref{fig:rosenberg2015}b. We conclude from our analysis that the resistive effects are unlikely to be the dominant force of the reconnection slowdown. We suggest that the fluid rebound effect could take place, preventing the current sheet from reaching the $d_i$ scale, thus sufficiently slowing down the reconnection. In other words, as we have an uncertainty in terms of the background density values, as well as the final curvature radius of the expanding bubbles, we may expect that we can get into the regions of no reconnection due to the rebound effect, as Figure 11 suggests.

{It is worth noting that we interpret results of an OMEGA EP experiment \cite{Rosenberg2015} while using the initial profiles obtained in OMEGA experiments \cite{RosenbergPoP2015,RosenbergNatComm} as our initial conditions. The main difference between profiles at OMEGA \& OMEGA EP is the temperature, due to different laser intensities. However, otherwise we expect the overall geometry to be very similar, as the laser spot sizes are very similar. In our simulations, we cover a wide range of Lundquist number values, which may be thought as covering a wide range of temperatures, which will therefore cover the OMEGA EP plasma parameters, too. }

\begin{figure*}
\includegraphics[width=0.99\linewidth]{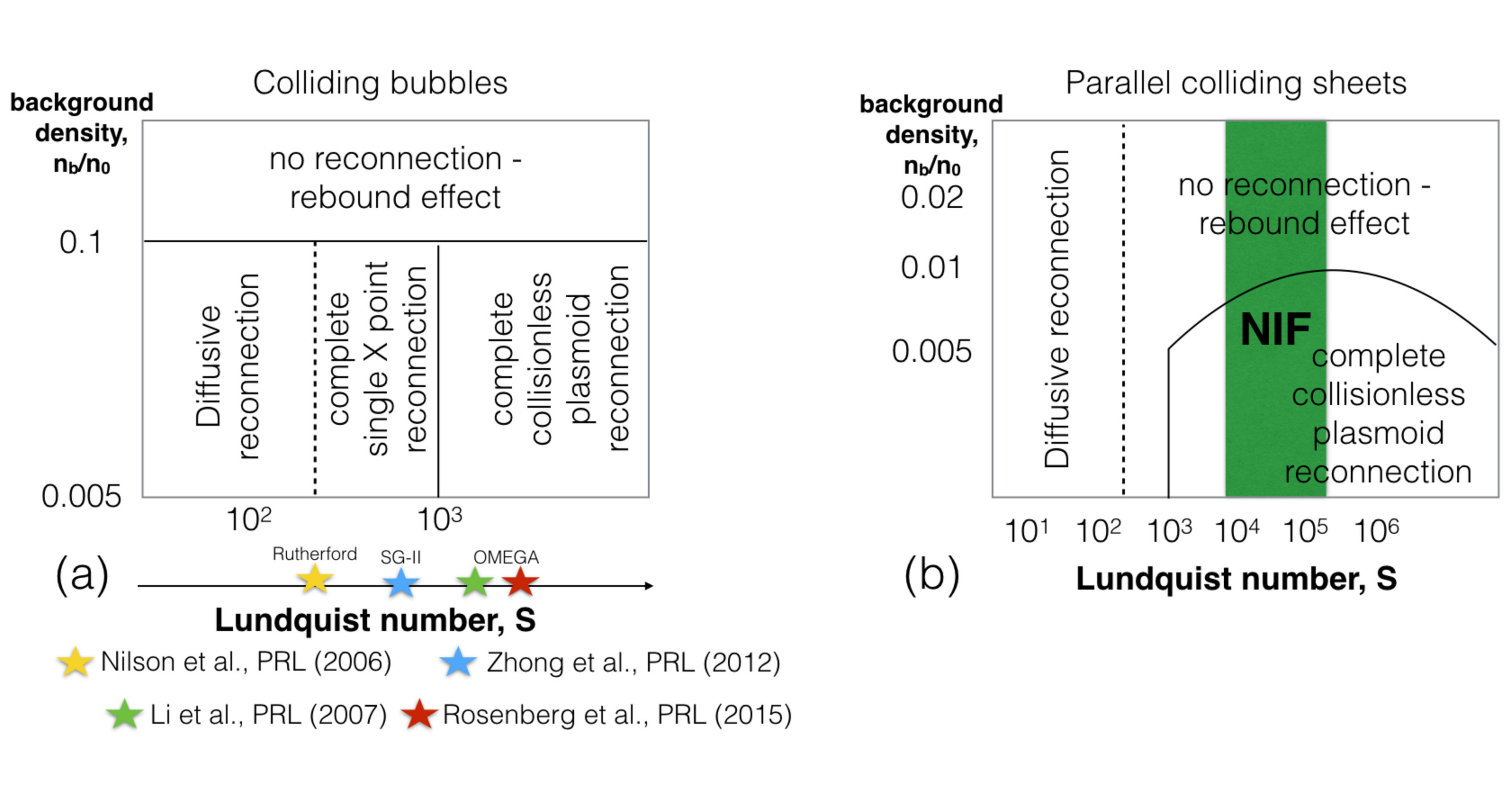}

\caption{(a) Parameter space related to recent experiments on magnetized bubble reconnection. The background density was not quantified in previous experiments, so we only show the location of these experiments on the Lundquist number axis. (b) Parameter space for parallel colliding sheets, also showing the region possible NIF experiments occupy (green stripe).}
\label{fig:summary}
\end{figure*}

\section{Discussion and Conclusions}
\label{sec:summary}

We have conducted a series of PIC simulations of driven reconnection experiments in order to assess the role of current sheet formation in the reconnection physics. We showed that the maximum reconnection rates are fast for all the simulation parameters considered -- all maximum rates are at least $0.1~ V_{A\ast}B_\ast$, which is natural for strongly driven reconnection (kinetic plasma $\beta_k = 8\pi (\rho V_{\rm inflow}^2/2)/B^2 \sim 10$ in our case). However, there is still some difference between the collisional ($S \leq 1000$) and collisionless ($S \gg 10^3$) cases, with collisionless reconnection being about a factor of 3 faster. The Sweet-Parker theory gives reconnection rates which are substantially lower than the values from the simulations, at least by a factor of ten. In the range of $S<10^3$, where we expect the rates to be in approximate agreement with the Sweet-Parker model ($d \psi/dt_{\rm SP} \propto S^{-1/2}$), our reconnection rates still exceed the SP ones. The reason is the strongly compressible plasma in the vicinity of the X point due to the strong supersonic inflow, which violates the plasma incompressibility assumption of the SP theory. The threshold for the transition to the faster rates is around $S \sim 10^3$. The threshold between the single X-point and multiple X-point, or plasmoid, reconnection lies in the same region of $S \sim 10^3$. It is shown that a very important criterion, $\delta/d_i < 1$, must be obtained to transition to a fast reconnection regime. The parallel collision case demonstrates the importance of fluid effects for the process of the current sheet formation. It was demonstrated that both collisionality and additional background plasma density may lead to the rebound effect, when the magnetized ribbons start moving away from each other, stalling the reconnection process. Figure~\ref{fig:summary} shows the ($S$,\,$n_b/n_0$) parameter region and marks the reconnection regimes that appear in our simulations.

It is instructive to compare the results of these simulations in the strongly driven HEDP regime to other PIC simulations which address driven reconnection in kinetic regimes, and transitions from collisional to collisionless reconnection.

First, Karimabadi et al. (2011) considers the problem of coalescence of two magnetic islands within collisionless 2D PIC simulations. Starting from the equilibrium, the merger process was initiated by the perturbation imposed on the system, and the influence of the island curvature radius on the merger efficiency was discussed. Even though the driving conditions in our simulations are dramatically different, our studies converge on multiple points. First, a flux pileup effect was observed both in Ref. \cite{Karimabadi2011} and in our simulations. Ref. \cite{Karimabadi2011} find a typical pileup factor of around two. However, we generally obtain higher pileup factors (for $S>10^4$) due to the more energetic inflows ($\beta_k \sim 10$) and, as a consequence, strongly compressed plasma in the current sheet (see Fig. 7, green line). Also consistent with the findings in  Ref. \cite{Karimabadi2011}, we demonstrate that for the resistive case that the parallel sheets geometry (curvature radius is formally infinite) shows much more effective bouncing in comparison to the bubble geometry (curvature radius is around $40\, d_i$) - the total reconnected flux for these cases was 60\% and 95\%, respectively). 

Second, Ref. \cite{DAUGTHON2009} studied the onset and transition from collisional to collisionless reconnection with PIC simulations that were initiated from Harris sheet. The initial stage of their simulation was in agreement with SP theory until the plasmoid instability takes place, after which SP theory breakes down and the thin current sheets are formed leading to fast reconnection. The reconnection process in our case is considerably different. We have a very wide current sheet in the form of two magnetised bubbles as the initial condition. Then, as the flow pushes magnetised stripes towards each other, the current sheet thins, and it is either disrupted by collisionless tearing (both parallel sheet and bubble geometry), reconnects via an X-point reconnection (bubble case), or completely rebounds (bubble and parallel sheet cases). This shows that the hydrodynamical time $L/C_s$ may compete with typical tearing time, thus influencing the outcome of reconnection. Nevertheless, the simulations both agree on a common point which is that the criteria for the fast reconnection onset is the same as in \cite{DAUGTHON2009} - $\delta/d_i<1$. It is instructive to note that this criterion was obtained earlier using Hall MHD simulations. See the review by Bhattacharjee et al. (2001) \cite{Bhattacharjee2001} and other references therein.

We also consider the results from \cite{Qiao2016}, which used a similar simulation setup to ours but which compared parallel and anti-parallel magnetic fields. Likewise, since our original simulations used an antiparallel magnetic field, we conducted additional simulations where the magnetic field was parallel. In \cite{Qiao2016}, it was shown that the plasma heating, out-of-plane magnetic field quadrupole geometry, and out-of-plane electric field cannot be considered as sufficient evidence for the occurance of magnetic reconnection. While our simulation geometry allows for plasmoid reconnection (unlike \cite{Qiao2016}), we reproduce these results. Electron jets are usually regarded as the sign of magnetic reconnection \cite{Rosenberg2015, JETS}, but our simulations show that their structure and typical velocities are nearly the same for both parallel and antiparallel cases, with maximum outflow velocitites up Mach 5, in agreement with the estimate from \cite{Rosenberg2015}. Recontructing the proton radiography pictures from our simulations, we conclude that it is indeed possible to detect the reconnection in the plasmoid regime - plasmoids appear as the zero-fluence circular regions in the overall homogenious picture. However, the proton radiography fluence maps are unable to differentiate between the collisional X-point reconnection and collisionless no-reconnection due to rebound. 

Thus, we have conducted a comprehensive sequence of 2.5D PIC simulations of driven reconnection in HEDP experiments. These simulations significantly expand the parameter space of previous simulations of reconnection in HEDP plasmas and illustrate a rich variety of behavior which can be potentially observed in experiments for comparison with theory.  We show that the plasma collisionality and background density, which in turn controls the compressibility of the plasma and width of the compressed current sheet, both play a role in determining the reconnection rates and evolution of the system. 
Similar to reconnection simulations in other geometries, we find again that that thinning the current sheet to the ion skin depth scale is essential for the onset of fast reconnection.
Sufficient background density can prevent thinning to this scale, which provides a possible explanation for recent experiments which show a stalled reconnection.
A novel feature of the HED system is the rapid evolution and formation of the current sheet, which is on the dynamic time scale $L/C_s$, and if reconnection cannot onset within this time scale, then the plasmas can hydrodynamically bounce, which completely shuts down the reconnection.   Obtaining such a rapid onset requires the formation of a thin current sheet where two-fluid effects to boost reconnection and tearing rates beyond pure resistive reconnection. Our results are relevant to the interpretation of experimental results obtained at OMEGA and future experiments at NIF.

\bigskip

\section{Acknowledgements}

{We would like to acknoledge R. P. J. Town and I. Igumenshchev for providing us with LASNEX and DRACO simulations data, respectively.} Simulations were conducted on the Titan supercomputer at the Oak Ridge Leadership Computing Facility at the Oak Ridge National Laboratory, supported by the Office of Science of the DOE under Contract No. DE-AC05-00OR22725.   This research was also supported by the DOE under Contracts No. DE-SC0008655 and No. DE-SC0016249.

\end{document}